\documentclass[12pt,epsf,amstex]{article}
\usepackage [dvips]{graphicx}
\usepackage{amsmath}
\usepackage{amssymb}
\usepackage{epsfig}
\addtocounter{secnumdepth}{1}
\begin{document}

\newcommand{\br}{\bar{r}}
\newcommand{\bbeta}{\bar{\beta}}
\newcommand{\bgamma}{\bar{\gamma}}
\newcommand{\bR}{{\bf{R}}}
\newcommand{\bS}{{\bf{S}}}
\newcommand{\half}{\frac{1}{2}}
\newcommand{\thalf}{\tfrac{1}{2}}
\newcommand{\prodm}{\prod_{m=1}^n}
\newcommand{\summ}{\sum_{m=1}^n}
\newcommand{\sumqno}{\sum_{q\neq 0}}
\newcommand{\tsum}{\Sigma}
\newcommand{\tPhi}{\bar{\Phi}}
\newcommand{\tPsi}{\bar{\Psi}}
\newcommand{\bsA}{\mathbf{A}}
\newcommand{\bsV}{\mathbf{V}}
\newcommand{\bsE}{\mathbf{E}}
\newcommand{\bsZ}{\hat{\mathbf{Z}}}
\newcommand{\bse}{\mbox{\bf{1}}}

\newcommand{\dd}{\mbox{d}}
\newcommand{\ee}{\mbox{e}}
\newcommand{\p}{\partial}

\newcommand{\bn}{\bar{n}}
\newcommand{\bN}{\bar{N}}
\newcommand{\cL}{\cal L}
\newcommand{\cW}{\cal W}

\newcommand{\la}{\langle}
\newcommand{\ra}{\rangle}

\newcommand{\beq}{\begin{equation}}
\newcommand{\eeq}{\end{equation}}
\newcommand{\bea}{\begin{eqnarray}}
\newcommand{\eea}{\end{eqnarray}}

\def\lsim{\:\raisebox{-0.5ex}{$\stackrel{\textstyle<}{\sim}$}\:}
\def\gsim{\:\raisebox{-0.5ex}{$\stackrel{\textstyle>}{\sim}$}\:}

\numberwithin{equation}{section}

\thispagestyle{empty}
\title{\Large {\bf New Monte Carlo method\\[3mm]
%for many-sided 
for planar Poisson-Voronoi cells\\
%of arbitrary sidedness\\ 
\phantom{xxx} }}

\author{{\bf H.\,J. Hilhorst}\\[5mm]
{\small Laboratoire de Physique Th\'eorique$^*$,
B\^atiment 210, Universit\'e de Paris-Sud}\\[-1mm]
{\small 91405 Orsay Cedex, France}\\}

\maketitle
\begin{small}
\begin{abstract}
\noindent 

By a new Monte Carlo algorithm we evaluate the sidedness probability
$p_n$ of a planar Poisson-Voronoi cell in the range $3 \leq n \leq 1600$.
The algorithm is developed on the basis of
earlier theoretical work; it exploits, in particular, 
the known asymptotic behavior of $p_n$ as $n\to\infty$. 
Our $p_n$ values all have between four and six significant digits. 
Accurate $n$ dependent averages, second moments, and variances 
are obtained for the cell area and the cell perimeter.
The numerical large $n$ behavior of these quantities is 
analyzed in terms of asymptotic power series in $n^{-1}$.
Snapshots are shown of typical occurrences of extremely rare events 
implicating cells of up to $n=1600$ sides embedded 
in an ordinary Poisson-Voronoi diagram. 
We reveal and discuss the
characteristic features of such many-sided cells and their
immediate environment. 
Their relevance for observable properties is stressed.\\

\end{abstract}
\end{small}
\vspace{35mm}

\noindent LPT -- ORSAY 06/95\\
{\small $^*$Research unit UMR 8627 of the Centre National de la
Recherche Scientifique}
\newpage

%%%%%%%%%%%%%%%%%%%%%%%%%%%%%%%%%%%%%%%%%%%%%%%%%%%%%%%%%%%%%%%%%%%%%%%%%%%%%
%%%%%%%%%%%%%%%%%%%%%%%%%%%%%%%%%%%%%%%%%%%%%%%%%%%%%%%%%%%%%%%%%%%%%%%%%%%%%

\section{Introduction} 
\label{secintroduction}
\vspace{5mm}

A Voronoi diagram
partitions space into convex cells constructed around a 
set of point-like `seeds' or `particles', 
in such a way that each point of space is in the cell of the particle
to which it is closest. 
When the particles are distributed randomly and uniformly, 
the partitioning is called a
Poisson-Voronoi diagram or a Random Voronoi Froth. 

Voronoi cells play a role throughout science and engineering
and are also of interest to mathematicians.
Applications include cellular structures that either 
arise spontaneously in nature (\textit{e.g.} in biological 
cellular structures, in soap froths, or in granular materials) or 
are employed as a tool of analysis (\textit{e.g.} 
to identify lattice defects in simulations of melting crystals). 
Many references are given in 
Ref.\,\cite{HJHart05} and in the encyclopedic monograph on tessellations 
by Okabe {\it et al.} \cite{Okabeetal00}.
 
The simplest Voronoi diagrams are of the Poisson type.
It is important, therefore, that the properties of Poisson-Voronoi 
diagrams be understood as well as possible. Here we pursue,
by means of a new Monte Carlo method, 
earlier investigations \cite{HJHart05,HJHletter05,HJHaboav06} 
on such diagrams in the Euclidean plane ${\mathbb R}^2$.
\vspace{2mm}

The most prominent
statistical property of the planar Poisson-Voronoi cell
is its `sidedness'. We denote by $p_n$ the probability that a cell
is $n$-sided, for arbitrary integer $n\geq3$.
Other properties of interest include 
the average area 
of an $n$-sided cell and the average length 
of its perimeter;  the statistics of the angles at the vertices; 
and correlations between neighboring cells. 
All these properties may be expressed
as multiple integrals on the particle positions
\cite{Meijering53,Okabeetal00}, but
only a few of them can be calculated explicitly. 
In particular, no simple closed form expression for $p_n$ is known.
An exact relation derived from
Euler's theorem ensures that the average sidedness 
$\overline{n} \equiv \sum_{n=3}^\infty np_n$
is equal to $\overline{n} =6$.

It is known numerically that $p_n$ 
peaks at $n=6$ and falls of rapidly for large $n$. 
Hayen and Quine \cite{HayenQuine00b} 
have numerically evaluated the integral for $p_3$ with high accuracy. 
For $n=4,5,\ldots\,$ the values of $p_n$ stem only from
Monte Carlo work. 
The most accurate reported values of $p_n$ are 
due to Calka \cite{Calka03b} for $4 \leq n \leq 7$
and to Brakke \cite{Brakke} for $8 \leq n \leq 16$.
One has $p_{16}\approx 10^{-8}$ and
the largest sidedness observed in simulations by conventional
algorithms is around $n=16$. 
Drouffe and Itzykson 
\cite{DI84,ID89}, as part of an effort to construct field theories on random
lattices, developed an improved algorithm by which they
estimated ${p}_n$ for $n$ up to $50$. Their results, however, have
error bars
that for $n\gsim 30 $  become of the same order as the $p_n$ 
themselves. 
Hence simulating many-sided Voronoi cells has remained a challenge.
\vspace{3mm}

The interest of investigating Voronoi cells for
asymptotically large $n$ was stressed by
Le Ca\"er and Delannay \cite{LeCaerDelannay93}.
Analytic knowledge of the large $n$ behavior of $p_n$,
apart from the insight that it provides, also
constrains the laws that describe the finite $n$ behavior
as observed in experiments and simulations.
An example of this interplay between the regimes of finite 
and of asymptotic $n$ is the theoretical explanation given in
Ref.\,\cite{HJHaboav06} of the failure of Aboav's law \cite{Aboav70}
for Poisson-Voronoi diagrams.

The analytic study of $p_n$ in the limit $n\to\infty$ was 
taken up in Refs.\,\cite{HJHletter05} and \cite{HJHart05}.
It was shown there, among many other things, that asymptotically 
\beq
p_n \simeq C p_n^{(0)}, \qquad n\to\infty,
\label{relpnaspt}
\eeq
with $C=0.344\,347...$ \cite{footnotee} and 
\beq
{p}_n^{(0)} = \frac{1}{4\pi^2}\,\frac{(8\pi^2)^n}{(2n)!}.
\label{resultpn0C}
\eeq
%%% [* 3631 num.value]
In the present work we exploit this asymptotic knowledge. Going beyond
Eq.\,(\ref{relpnaspt}) we write an equality that is exact for all $n$ rather
than merely asymptotic, namely
\beq
p_n = C_n p_n^{(0)},
\label{defCn}
\eeq
whence necessarily $\lim_{n\to\infty} C_n = C$.
In this work we focus on $C_n$ and
show that it can be expressed as an average 
\beq
C_n\,=\,\la\Theta\ee^{-{\mathbb V}}\ra,
\label{relCnexpmV}
\eeq
where ${\mathbb V}$ is a known expression
in the angular variables that describe the $n$-sided cell,
and $\Theta$ is an indicator (\textit{i.e.,} 
equal to $0$ or to $1$) imposing
a geometric constraint on the set of angles.
We will determine the prefactor $C_n$ in (\ref{defCn})
by Monte Carlo evaluation of the right hand side of Eq.\,(\ref{relCnexpmV})
for finite $n=3,4,\ldots$.
The Monte Carlo algorithm is new for this problem.
Whereas all previously used methods become rapidly inefficient
with increasing $n$, the performance of the algorithm presented
here is, very roughly, independent of $n$. This makes it possible, in
particular, to explore the structure of Voronoi cells in the
hitherto inaccessible large-$n$ regime.

The remaining sections of this paper are the following. 
In Sec.\,\ref{secMonteCarlo} the algorithm is described. 
In Sec.\,\ref{secresults} results are presented and discussed
for the sidedness probability $p_n$ as well as for the averages and second
moments of the cell perimeter and cell area. 
The asymptotic large-$n$ behavior of these quantities is analyzed numerically.
In Sec.\,\ref{secshapes} we present and discuss characteristic pictures of
many-sided Voronoi cells in an environment of ordinary cells.
In Sec.\,\ref{secconclusion} we summarize our results.

The algorithm requires the explicit expressions for ${\mathbb V}$
and $\Theta$
in Eq.\,(\ref{relCnexpmV}). Finding these
is a matter of considerable technical complexity; 
it is based on results of Ref.\,\cite{HJHart05} and is the subject of
Appendices \ref{apptheory} and \ref{appeqnG0}.

%%%%%%%%%%%%%%%%%%%%%%%%%%%%%%%%%%%%%%%%%%%%%%%%%%%%%%%%%%%%%%%%%%%%%%%%%%%

\section{Monte Carlo algorithm}
\label{secMonteCarlo}

\subsection{Context}
\label{secintroMC}

Monte Carlo methods for generating Voronoi cells of Poisson
distributed particles are discussed in the monograph by Okabe {\it et al.} 
\cite{Okabeetal00}.
One class of methods simply determines $p_n$ as the relative frequency of
occurrence of $n$-sided cells. 
But since $p_n$ decreases to zero faster than exponentially
for $n \gsim 12$, 
the statistical precision goes down accordingly.
With such methods
it is hardly possible to accumulate sufficient statistics for 
even single-digit
precision as soon as $n \approx 16$.

Another class of methods 
generates cells for a value of $n$ fixed in
advance. The first to have done so seem to have been Drouffe and Itzykson 
\cite{DI84}. The method employed by Calka \cite{Calka03b} is also in this
class. These methods face the problem of attrition:
a Monte Carlo generated geometrical object, in order to represent a valid
$n$-sided cell, must satisfy certain geometrical constraints.
The probability that an attempted
generation satisfy the constraints again decreases rapidly with growing $n$.

The present algorithm, which also fixes $n$ in advance,
completely solves the problem of attrition: 
the geometric constraints are satisfied with a probability that
tends to unity when $n \to \infty$. In order to arrange things this way,
a certain amount of rather technical rewriting of the initial problem is
necessary. We have confined this rewriting to the Appendices. 
If one accepts its results, the method is easy to apply.

%%%%%%%%%%%%%%%%%%%%%%%%%%%%%%%%%%%%%%%%%%%%%%%%%%%%%%%%%%%%%%%%%%%%%%%%%%%%%

\subsection{Angular variables}
\label{secMCintroduction}

\begin{figure}
\begin{center}
\scalebox{.60}
{\includegraphics{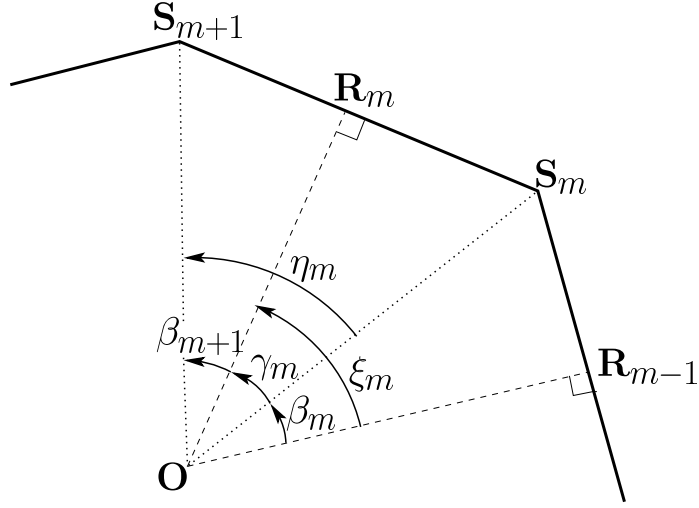}}
\end{center}
\caption{{\small Heavy line: the perimeter of the Voronoi cell around a
particle in the origin $\mbox{\bf O}$.
Dashed and dotted lines connect the origin to the
midpoints $\bR_m$ and vertices $\bS_m$, 
respectively. 
The particles of the neighboring cells are located at $2\bR_1,\ldots,2\bR_n$. 
Right angles have been marked.
The figure defines the sets of 
angles $\xi_m$, $\eta_m$, $\beta_m$, and $\gamma_m$.}}
\label{figdefangles}
\end{figure}

An $n$-sided Voronoi cell around a particle in the origin,
as shown in Fig.\,\ref{figdefangles},  
is specified completely by its
$n$ vertex vectors $\bS_1,\bS_2,\ldots,\bS_n$. It may be specified
alternatively by its $n$ mid-point vectors, \textit{i.e.} the
projections $\bR_1,\bR_2,\ldots,\bR_n$
of the origin onto the sides.
The explicit expression \cite{DI84,Calka03a,Calka03b,HJHart05} 
for $p_n$ as a multiple integral
on the $\bR_m$ is given in Appendix \ref{apptheory}.
It has not, however, been possible to evaluate this integral
analytically.
By choosing other sets of variables of integration one may recast
the original integral in numerous different forms, none of which is simple.
For our purpose it is essential to use the angular variables 
that we will define now. 
%At the end of this subsection we will indicate how
%$\Theta$ and ${\mathbb V}$ in Eq.\,(\ref{relCnexpmV})
%depend on them.

Let $\Phi_m$ and $\Psi_m$ be the polar angles of $\bR_m$ and $\bS_m$,
respectively. Other angles
relevant for this problem are defined in Fig.\,\ref{figdefangles}.
The $\eta_m=\Psi_{m+1}-\Psi_m$ are the angles between two consecutive vertex
vectors and the  
$\xi_m=\Phi_m-\Phi_{m-1}$ those between two consecutive vertex vectors; 
$n$-periodicity in the index $m$ is understood.
For fixed sets 
$\xi=\{\xi_m\}$ and $\eta=\{\eta_m\}$ one may still jointly rotate
the set of vertex vectors with respect to the set of mid-point vectors:
this modifies only the relative angles $\beta_m$ and $\gamma_m$ 
between the two sets. 
We may select any one of these relative angles and call it `the' angle of
rotation, since it will determine all others; we will select $\beta_1$.
When for a generic $\beta_1$ we draw
the cell boundary by clockwise constructing its successive segments, 
then after a turn of $2\pi$ it appears not close onto
itself but to spiral. 
A `no-spiral condition' must therefore determine
the appropriate value of the
angle of rotation $\beta_1$ for which the cell boundary closes
and which we will denote by $\beta_1=\beta_*(\xi,\eta)$. 
This condition reads
\beq
G(\xi,\eta;\beta_*)=0
\label{nospiralcond}
\eeq
where
\beq
\ee^{2\pi G} = \prod_{m=1}^n \frac{\cos\gamma_m}{\cos\beta_m}\,.
\label{defG}
\eeq
One may note that Eq.\,(\ref{defG}) involves the
$\beta_m$ and $\gamma_m$ that
are themselves determined by the solution $\beta_1=\beta_*$ of
(\ref{nospiralcond}).
For an arbitrary pair $(\xi,\eta)$ there need not exist a
solution to Eq.\,(\ref{nospiralcond}).
In Appendix \ref{appeqnG0} we show that 
it has a solution, which moreover is unique, 
if and only if
\beq
\max_{1\leq m\leq n} \Big[ \sum_{\ell=1}^{m-1}(\xi_\ell-\eta_\ell)\,+\,
\xi_m \Big]\, -\,
\min_{1\leq m\leq n} \Big[ \sum_{\ell=1}^{m-1}(\xi_\ell-\eta_\ell) \Big]\, 
<\, \pi,
\label{nospiraldomain}
\eeq
which is a criterion expressed entirely in terms of the supposedly given
sets $\xi$ and $\eta$.

After these preliminaries we return to (\ref{relCnexpmV}).
The symbol $\Theta$ in that expression denotes the indicator function
of the domain in $(\xi,\eta)$ space
where (\ref{nospiraldomain}) is satisfied.
Finally, the `interaction' ${\mathbb V}$ in (\ref{relCnexpmV})
is given explicitly in terms of the angular variables 
in Appendix \ref{apptheory} through a sequence of
definitions, Eqs.\,(\ref{defmathbbV}) and 
(\ref{defTm})-(\ref{exprVbetagamma}), that we will not display here.

%%%%%%%%%%%%%%%%%%%%%%%%%%%%%%%%
%%%%%%%%%%%%%%%%%%%%%%%%%%%%%%%%
%%%%%%%%%%%%%%%%%%%%%%%%%%%%%%%%
%%%%%%%%%%%%%%%%%%%%%%%%%%%%%%%%

%%%%%%%%%%%%%%%%%%%%%%%%%%%%%%%%%%%%%%%%%%%%%%%%%%%%%%%%%%%%%%%%%%%%%%%%%%%%%

\subsection{Algorithm for determining $p_n$}
\label{secpn}

The sidedness probability $p_n$ is given by 
Eqs.\,(\ref{resultpn0C})-(\ref{relCnexpmV}). We determine it numerically
by evaluating $\la \Theta\ee^{-{\mathbb V}} \ra$ as follows.
We fix the sidedness $n$,
after which the simulation proceeds as follows.

(i) We draw $n-1$ random numbers uniformly distributed on $[0,1]$
and order them. After multiplication by $2\pi$ this gives \cite{footnotec}
$0 < \tPsi_1 < \tPsi_2 < \ldots < \tPsi_{n-1}< 2\pi$.
We set $\tPsi_n=2\pi$ and choose
\bea
\eta_m &=& \tPsi_{m+1}-\tPsi_{m}\,, \qquad m=1,\ldots,n-1,\nonumber\\
\eta_n &=& \tPsi_{1}.
\label{defetam}
\eea
We next draw $2n-1$ random numbers, order them, and discard
those of odd rank so that only $n-1$ are left. After multiplication by
$2\pi$ this gives 
$0 < \tPhi_1 < \tPhi_2 < \ldots < \tPhi_{n-1} < 2\pi$. We set
$\tPhi_0=0$ and choose
\bea
\xi_m &=& \tPhi_m - \tPhi_{m-1}\,, \qquad m=1,\ldots,n-1,\nonumber\\
\xi_n &=& 2\pi - \tPhi_{n-1}\,.
\label{defxim}
\eea

(ii) We check if the pair of sets 
$(\xi,\eta)$ thus obtained
satisfies Eq.\,(\ref{nospiraldomain}).
If so, then we know that there exists a $\beta_*(\xi,\eta)$
which may be determined from Eq.\,(\ref{nospiralcond}), hence $\Theta=1$ and
we proceed with (iii). If not, then
it is impossible to satisfy Eq.\,(\ref{nospiralcond}), we have $\Theta=0$, 
and the attempt to generate an $n$-sided cell fails. 
We increase the attempt counter by one unit and return to (i). 

(iii) 
We solve $\beta_*(\xi,\eta)$ from Eq.\,(\ref{nospiralcond}) 
by a numerical iteration procedure 
which also yields the derivative $G'(\xi,\eta;\beta_*)$ needed in the
next step.

(iv) We calculate the weight $\exp(-{\mathbb V})$
according to Eqs.\,(\ref{defmathbbV}) and 
(\ref{defTm})-(\ref{exprVbetagamma}) of Appendix \ref{apptheory}
and add it to the accumulated weight. We increase the attempt counter by one
unit and return to (i).

(v) In the end the total accumulated weight is divided by the total
number of attempts, including those that failed. The result is an estimate for
$p_n$. 

We remark that 
the successive cells generated by this procedure
are all statistically independent.

%%%%%%%%%%%%%%%%%%%%%%%%%%%%%%%%%%%%%%%%%%%%%%%%%%%%%%%%%%%%%%%%%%%%%%%%%%%%%

\subsection{Algorithm for $n$ dependent averages}
\label{secproperties}

The simulation method described above
allows us to study arbitrary cell properties $F(\bR_1,\ldots,\bR_n)$. 
Writing $\la F\ra_n$ for the average of $F$ subject to a given
sidedness $n$ we have
\beq
\la F \ra_n =\frac{\la I_F \Theta\ee^{-{\mathbb V}}\ra}
                {\la\Theta\ee^{-{\mathbb V}}\ra}.
\label{defFavn}
\eeq
Here the numerator, which generalizes (\ref{relCnexpmV}),
has an insertion $I_F$ that derives from $F$ by a radial integration.
We recall that the average $\la\ldots\ra$, defined in (\ref{pnfin}), 
applies to quantities that depend exclusively on the angular variables. 
To find $I_F$ from $F$, we
let $R_{\rm av}$ denote the average of the $R_m$.
Upon setting $R_m=R_{\rm av}\rho_m$
we may express the $\rho_m$
entirely in terms of the angular variables (see Appendix \ref{apptheory}).
Then, if $F$ is of dimension $d_{_F}$, it
may be factorized into a radial and an angular part according to
\beq
F(\bR_1,\ldots,\bR_n) = (R_{\rm av}^2/4\lambda)^{d_{_F}/2} \hat{F}(\xi,\eta),
\label{exprF}
\eeq
where we show explicitly the areal particle density $\lambda$ which had
heretofore  
been scaled away \cite{footnotef}. 
When (\ref{exprF}) is integrated over the radial scale $R_{\rm av}$,  
an extra factor appears as compared to the same operation for $p_n$
and we find
\beq
I_F = \frac{\Gamma(n+\tfrac{1}{2}d_{_F})}{\Gamma(n)} W^{-d_{_F}/2} \hat{F}\,,
\label{defIF}
\eeq
where $\Gamma$ denotes the Gamma function
and where we abbreviated
\beq
W = 4\lambda\pi(1+n^{-1}V)
\label{defW}
\eeq
with $V$ given by (\ref{exprVbetagamma}).

We will limit ourselves to considering the first and second moments of two 
quantities that are frequently encountered in applications and that
have therefore been the subject of much earlier work, \textit{viz.}
the cell perimeter $P$ and the cell area $A$. These are explicitly given by
\beq
P = R_{\rm av} (4\lambda)^{-1/2} \hat{F}_1, \qquad 
A = R_{\rm av}^2 (4\lambda)^{-1} \hat{F}_2\,,
\label{exprPA}
\eeq
with the angular factors
\beq
\hat{F}_k = \frac{1}{k}
\sum_{m=1}^n \rho_m^k (\tan\gamma_m + \tan\beta_{m+1}), \qquad k=1,2.
\label{defF12}
\eeq
Setting successively $F=P, P^2, A, A^2$ we find that
the corresponding insertions in the numerator of Eq.\,(\ref{defFavn}) are
\bea
I_P\,\, &=& [\Gamma(n+\tfrac{1}{2})/\Gamma(n)]
                               W^{-1/2} \hat{F}_1\,,   \nonumber\\
I_{P^2} &=& n                  W^{-1}   \hat{F}_1^2\,, \nonumber\\
I_A\,\, &=& \tfrac{1}{2}n      W^{-1}   \hat{F}_2\,,   \nonumber\\
I_{A^2} &=& \tfrac{1}{4}n(n+1) W^{-2}   \hat{F}_2^2\,.
\label{exprinsertions}
\eea
The simulation steps for finding the numerator of Eq.\,(\ref{defFavn}) 
are the same as for $p_n$ except that (iv) and (v) are replaced with
(iv$^\prime$) and (v$^\prime$) given below.

(iv$^\prime$) We multiply the insertion $I_F$ of the quantity $F$ of interest
by the weight $\exp(-{\mathbb V})$ and 
accumulate the value thus obtained.

(v$^\prime$) In the end the total accumulated value is divided by the total
number of attempts and by the estimate obtained for $p_n$.
This provides an estimate for $\langle F \rangle_n$.
The numerical data shown will all be for $\lambda=1$.

%%%%%%%%%%%%%%%%%%%%%%%%%%%%%%%%%%%%%%%%%%%%%%%%%%%%%%%%%%%%%%%%%%%%%%%%%%%%%

\section{Results and discussion}
\label{secresults}

%%%%%%%%%%%%%%%%%%%%%%%%%%%%%%%%%%%%%%%%%%%%%%%%%%%%%%%%%%%%%%%%%%%%%%%%%%%%%

\subsection{The distribution of ${\mathbb V}$ and the indicator $\Theta$}
\label{secdistrmathbbV}

Before discussing our results for the sidedness probability $p_n$
we briefly consider the quantities
${\mathbb V}$ and $\Theta$ that via
(\ref{defCn}) and (\ref{relCnexpmV})
enter into its definition. Let $P({\mathbb V})$ denote
the probability distribution of ${\mathbb V}$ and 
$\phi_n\equiv\la\Theta\ra$ the probability for an attempted cell generation
to be successful. In terms of these we may rewrite
(\ref{relCnexpmV}) as  
\beq
C_n = \phi_n \int\! \dd{\mathbb V}\, P({\mathbb V})\ee^{-{\mathbb V}},
\label{relCnphiV}
\eeq
which exhibits the important intermediate role of $P({\mathbb V})$
and $\phi_n$.

\begin{figure}
\begin{center}
\scalebox{.60}
{\includegraphics{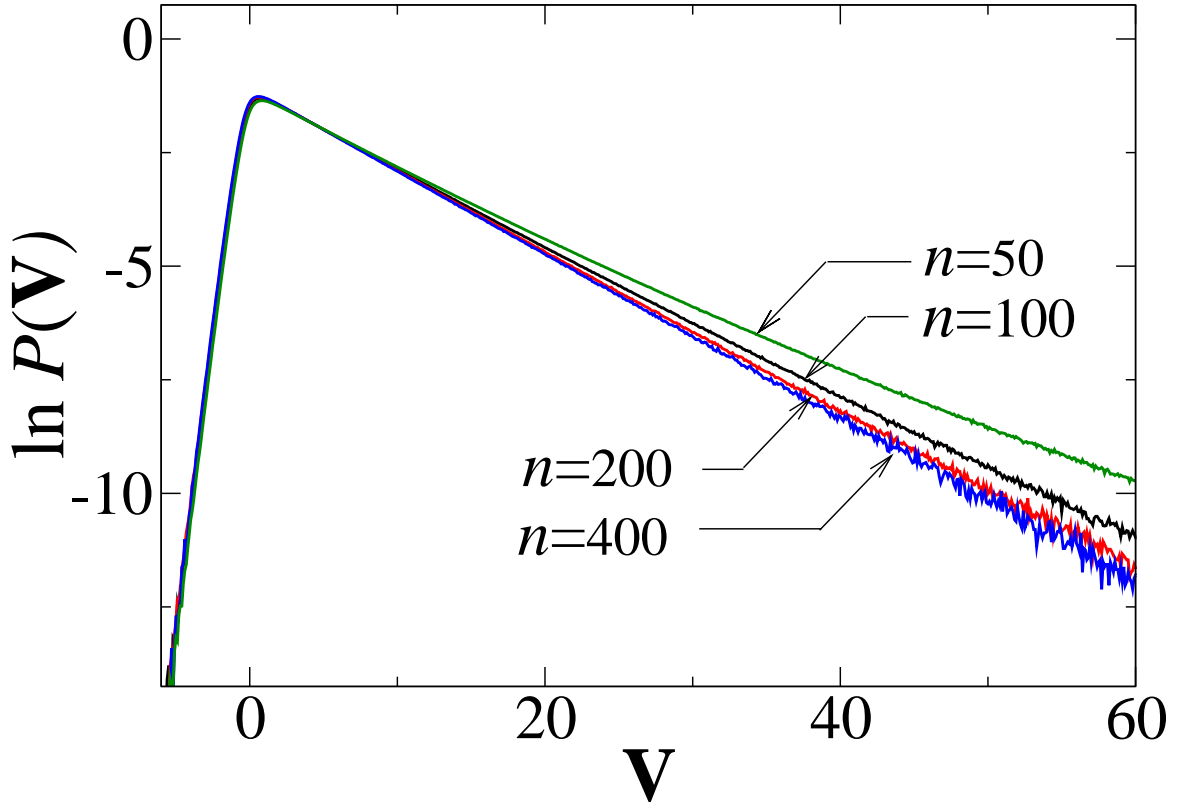}}
\end{center}
\caption{\small Logarithm of the probability distribution 
$P({\mathbb V})$ of ${\mathbb V}$ [see Eq.\,(\ref{relCnphiV})]
for four different values of $n$,
showing convergence to a limit distribution for $n=\infty$.
}
\label{figlnPV}
\end{figure}

In order to show what $P({\mathbb V})$ looks like, we have plotted 
its logarithm 
in Fig.\,\ref{figlnPV} for $n=50,\, 100,\, 200,$ and $400$. 
The curves clearly
demonstrate that for $n\to\infty$ there is convergence to a limit.
For ${\mathbb V}\to\pm\infty$
the limit distribution appears to decay exponentially, 
$P({\mathbb V}) \sim \exp(-\kappa_{\pm}|{\mathbb V}|)$,
but with very different decay constants: we obtain 
$\kappa_+=0.185\pm0.05$ from a fit in the range $3\leq{\mathbb V}\leq 30$
followed by extrapolation to $n=\infty$, and
$\kappa_-=2.47\pm0.02$ from a fit in the range $-3\leq{\mathbb V}\leq-1.5$.
This large ${\mathbb V}$ behavior has not yet been explained theoretically.

The $\Theta$ function in (\ref{relCnexpmV}) imposes 
constraint (\ref{nospiraldomain}) and
is at the origin of the failed generation attempts.
Whereas these do not contribute to $p_n$
in step (iv) of the algorithm above,
Eq.\,(\ref{relCnphiV}) shows that via $\la\Theta\ra=\phi_n$ 
they do enter into the determination of its normalization.
In the last column of Table \ref{tablepn} we list the fractions $\phi_n$
of successful attempts as determined from the simulation.
Although $\phi_n$ is equal only to $\phi_3=0.058$ for $n=3$, 
it turns out to rise rapidly with 
$n$, is already as high as  $p_{10}=0.8$ for $n=10$,
and tends to unity for $n\to\infty$. That is, attrition disappears in the
large $n$ limit.
\vspace{3mm}

This brings out the two key steps 
that are responsible for the success of the present algorithm:
(i) the limit distribution of $P({\mathbb V})$ 
has become $n$ independent 
since we extracted from the initial expression for $p_n$ 
the appropriate $n$ dependent prefactor $p_n^{(0)}$ given in 
(\ref{resultpn0C}); and
(ii) attrition disappears for large $n$
because of our choice of
the angles $(\xi,\eta)$ as the variables of integration.

%%%%%%%%%%%%%%%%%%%%%%%%%%%%%%%%
%%%%%%%%%%%%%%%%%%%%%%%%%%%%%%%%
%%%%%%%%%%%%%%%%%%%%%%%%%%%%%%%%
%%%%%%%%%%%%%%%%%%%%%%%%%%%%%%%%

%%%%%%%%%%%%%%%%%%%%%%%%%%%%%%%%%%%%%%%%%%%%%%%%%%%%%%%%%%%%%%%%%%%%%%%%%%%%%

\subsection{Sidedness probability $p_n$}
\label{secresultspn}

In Table \ref{tablepn} 
we present the results for the sidedness 
probability $p_n$ for $n$ in the range between $n=3$ and $n=1600$.
They are based on a number
$N_n$ of generation attempts 
given in the fourth column of that table.

The second column of Table \ref{tablepn} shows the best results for $p_n$ 
found in the literature for each value of $n$.
The $p_3$ value was obtained by numerical integration, the other $p_n$
are Monte Carlo results. 
For $p_4,\ldots,p_7$ the statistical error is in the last decimal;
for higher $n$ the standard deviations are indicated.

Our own results for $p_n$,
given in the third column of Table \ref{tablepn},
are accurate up to absolute errors 
of order less than $10^{-5}$.
Standard deviations were calculated by subdividing the data into 
twenty or more groups and considering the dispersion of the group averages.
We will now discuss these results as a function of $n$. 

{\it Case $n=3$. --\,\, } The probability $p_3$ for a cell to be 
three-sided is
the only one that has been evaluated by numerical integration.
This was done by Hayen and Quine \cite{HayenQuine00b}, who
reduced the original integral to a four-dimensional one.
They present a 12-digit result
of which the second column of Table 
\ref{tablepn}
shows only the first seven significant decimals.
For $n=3$ we performed an especially long run with the purpose of
testing our Monte Carlo method and checking the result of 
Ref.\,\cite{HayenQuine00b}. As shown in Table 
\ref{tablepn},
our method reproduces six significant digits of Hayen and Quine's result
and leaves their value within our error bars.

For all $n>3$ the literature results are based on Monte Carlo evaluation. 

{\it Cases $n=4$ through  $n=7$. --\,\, }
The most accurate literature results in this intermediate
regime are due to Calka \cite{Calka03b}, whose algorithm
like ours fixed $n$ in advance.
Our results are fully compatible with those of Ref.\,\cite{Calka03b}.

{\it Cases $n=8$ through $n=15$. --\,\,}
Monte Carlo results obtained in the 1980's by Brakke \cite{Brakke}
for $3 \leq n \leq 16$ stayed for a long
while the most accurate ones that were available.
Our simulations confirm all of Brakke's results.
Beyond $n \approx 10$ the accuracy of the
Monte Carlo algorithm of Ref.\,\cite{Brakke} 
rapidly goes down with increasing $n$,
and for $n=16$ its error bars are as large as the result itself.
This effect is simply due to the low relative frequency
of cells of so many sides, 
the number $n$ not being fixed in this method.
By contrast, the accuracy of our method, for a fixed amount
of computer time invested per value of $n$, stays roughly constant.

{\it Case $n \geq 16$. --\,\,}
Drouffe and Itzykson \cite{DI84}
developed a more powerful simulation method aimed at
simulating cells of larger sidedness. In their method $n$ is again
fixed in advance. 
Their accuracy amounts to roughly a single significant digit in the
regime $16 \leq n \lsim 25$; for $n \gsim 25$
the error becomes again of the order of $p_n$ itself. This
error increase is due to attrition, {\it i.e.,} 
an increasing rejection rate of configurations 
that are generated but do not satisfy the required geometrical constraints.
From our data it appears that for $n \gsim 25$ the work of
Ref.\,\cite{DI84} misses the true values by an ever larger factor
and that only their logarithmic order of
magnitude is right \cite{footnoteg}. 
Again, the method of the present work maintains an error in the fifth digit,
{\it i.e.,} a relative error not larger than $10^{-4}$.

{\it Case of extremely large  $n$. --\,\,}
The range of $n$ from 50 to 1600 had so far been unexplored
territory.  In this very large $n$ regime
attrition is negligible
and, for a constant calculational effort per value of $n$, 
the method keeps producing results with an error only in the fourth 
significant digit.
The probabilities $p_n$ are extremely small. 
Numerically we could easily handle such small numbers  
by first factorizing out $p_n^{(0)}$ of which we
computed and stored only the logarithm. 
As discussed in subsection \ref{secdistrmathbbV},
the remaining factor 
$C_n=\la\Theta\ee^{-{\mathbb V}}\ra$ has a finite distribution 
and hence causes no underflow problems. 

Generating the values of such `unphysically' small probabilities
is much more than a mere technical achievement.
First, it provides another 
check that the program works correctly; indeed we find
that for $n\to\infty$ the ratio $p_n/Cp_n^{(0)}=C_n/C$
tends to unity as it should.
Second, it gives access to the large $n$
expansion of $p_n$ to be discussed in subsection \ref{secasymptotic}.
Thirdly and most importantly, 
values of $n$ this large are required to 
see the separation of length scales that occurs in the many-sided cell; 
this is the subject of section \ref{secshapes}.

{\it Sum rules. --\,\,}
The $p_n$ should obey certain sum rules. 
Upon summing the $p_n$ of Table \ref{tablepn}
and writing $\overline{X_n}=\sum_{n=3}^\infty X_n p_n$ we find
\bea
\sum_{n=3}^\infty p_n   &=& 1.000\,010 (15), \nonumber\\
\overline{n} &=& 6.000\,1 (1), \nonumber\\
\overline{n^2} &=& 37.781\,6 (7),\nonumber\\
\mu\equiv\overline{n^2}-{\overline{n}}^2 &=& 1.780\,4,
\label{sumrules}
\eea
with an error in $\mu$ at most equal to $\pm 0.0015$
but probably smaller due to partial cancellation 
of the errors in ${\overline{n}}^2$ and $\overline{n^2}$.
The first and second relations of (\ref{sumrules}) may
be compared to the exactly known values $1$ and $6$, respectively.
The second moment $\overline{n^2}$ has an exact expression as a double
integral \cite{Finch05}, which when evaluated numerically gives
$\overline{n^2}=37.780\,811...$. Hence $\mu=1.780\,811...$ numerically
exactly. We therefore see that, when their error bars are taken into account, 
our Monte Carlo data are in excellent agreement with these sum rules.
\vspace{3mm}

{\it Conclusion. --\,\,} 
The general conclusion of this subsection is that
for low $n$ (say $n \lsim 8$) our method is probably as good as 
several of the existing ones. If we did slightly outperform them 
in that small $n$ regime,
that was only because of the length of our runs.
However, for larger $N$ (say $n \gsim 8$) our method 
has a decisive advantage over the existing ones.

%%%%%%%%%%%%%%%%%%%%%%%%%%%%%%%%%%%%%%%%%%%%%%%%%%%%%%%%%%%%%%%%%%%%%%%%%%%%%%

\subsection{Asymptotic behavior of $p_n$}
\label{secasymptotic}

On the basis of the numerical data we will now discuss the
asymptotic behavior of $p_n$ for $n \to \infty$.
Analytically it is known that $p_n=C_np_n^{(0)}$ with $p_n^{(0)}$ 
given by Eq.\,(\ref{resultpn0C}) and where the correction factor $C_n$ 
may be obtained by a series expansion
that classifies contributions according to their power in $n^{-1/2}$.
%In Refs.\,\cite{HJHletter05,HJHart05}  
%the leading order corrections were therefore written as ${\cal O}(n^{-1/2})$.
On that basis
Ref.\,\cite{HJHletter05} fitted the limited $p_n$ data available at
that time (essentially $n\lsim 30$) 
by a correction term proportional to $n^{-1/2}$.
It remained possible, however, that the coefficient of the
$n^{-1/2}$ term would cancel,
and indeed Drouffe and Itzykson \cite{DI84} had hypothesized earlier
that the leading correction was of order $n^{-1}$. 

The numerical data of this work now indicate unambiguously
that the series is actually one in powers of $n^{-1}$,
\beq
p_n=\frac{C}{4\pi^2}\,\frac{(8\pi^2)^n}{(2n)!}\Big[\,1\,-\,\frac{e_1}{n}
\,+\,\frac{e_2}{n^2}\,-\,\frac{e_3}{n^3}\,+\,\ldots\,\Big],
\label{asptexppn}
\eeq
where $e_1,e_2,\ldots,$ are numerical coefficients. 

\begin{figure}
\begin{center}
\scalebox{.60}
{\includegraphics{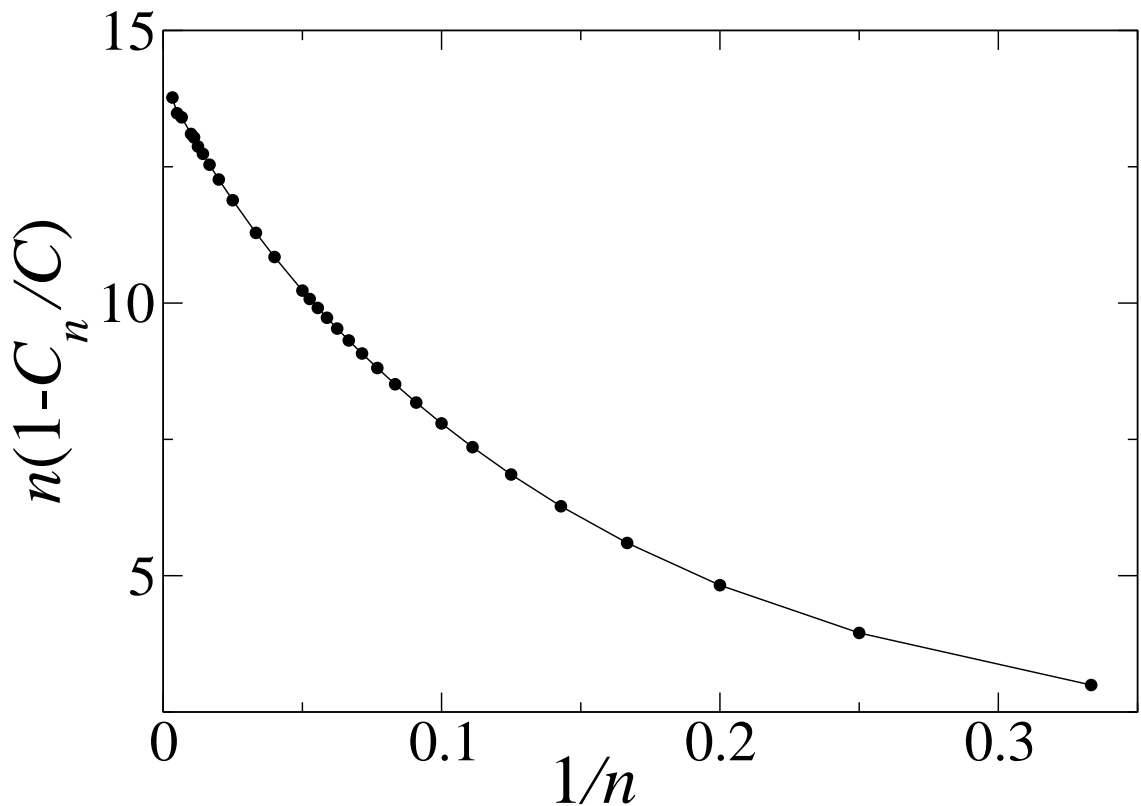}}
\end{center}
\caption{{\small To study the asymptotic large $n$ behavior of the sidedness
    probability $p_n=C_np_n^{(0)}$ 
    we plot the quantity $n(1-C_n/C)$ where $C=\lim_{n\to\infty}C_n$
    [see Eqs.\,(\ref{resultpn0C}) and (\ref{defCn})].  
The solid line connects the data points.
Data shown are in the range $3 \leq n \leq 300$.
The largest error bars occur for small $1/n$ and are of the order
of the data symbols. 
The intercept of the curve with the vertical axis
is the coefficient $e_1$ of the leading correction term in the expansion of
Eq.\,(\ref{asptexppn}). 
}}
\label{figcncfrot}
\end{figure}

The factor in square brackets in Eq.\,(\ref{asptexppn}) is equal to $C_n/C$,
which for $n\to\infty$ is known to tend to unity.
In Fig.\,\ref{figcncfrot}, in order to find the corrections to the leading
order term in (\ref{asptexppn}), we have plotted 
$n(1-C_n/C) = e_1-e_2n^{-1}+\ldots$ against $n^{-1}$. 
This figure shows that
the intercept with the vertical axis is located at
$e_1=14.00\pm 0.05$. We may now proceed by subtracting this estimated
value of $e_1$ from the curve of Fig.\,\ref{figcncfrot},
multiply it again by $n$, and 
look for a new intercept with the vertical axis
which, if it is well-defined, is equal to $-e_2$.
Upon iterating until the statistical
errors obscure a well-defined 
intercept we obtained in this way
estimates for the first few $e_i$. The uncertainties  
increase with the index $i$. We found 
\beq
e_1 =  14.00 \pm 0.05, \quad               
e_2 =  94 \pm 4, \quad                            
e_3 = 375 \pm 80,
\label{valuesei}
\eeq
in which the errors are correlated: the values deviate together
upward or downward.
The important conclusion is that $p_n$ has a series expansion in powers
of $n^{-1}$. The cancellation of the half-integer powers in the expansion of
Ref.\,\cite{HJHart05} is no doubt due to a symmetry in the theory
that still remains to be identified. 

In a final remark we wish to
stress that finding this asymptotic expansion is different from finding
a `best fit', which we do not attempt here.
The curve of Fig.\,\ref{figcncfrot} is close 
to the sum of a constant and an exponential in $n^{-1}$, 
but we have no reason to believe that there exists a simple 
analytical expression that fits all data within their error bars. 

%%%%%%%%%%%%%%%%%%%%%%%%%%%%%%%%%%%%%%%%%%%%%%%%%%%%%%%%%%%%%%%%%%%%%%%%%%%%%

%%%%%%%%%%%%%%%%%%%%%%%%%%%%%%%%
%%%%%%%%%%%%%%%%%%%%%%%%%%%%%%%&

\begin{table}
\begin{footnotesize}

\begin{center}
\renewcommand{\arraystretch}{1.1}
\begin{tabular}{||r|r|r r r||}
\hline
\multicolumn{1}{||r|}{}&
\multicolumn{1}{|c|}{ Literature \cite{HayenQuine00b}-\cite{DI84} }&
\multicolumn{3}{|c||}{ This work }\\

\multicolumn{1}{||r|}{$n$}&
\multicolumn{1}{ |c|}{$p_n$}&
\multicolumn{1}{ |c }{$p_n$}&
\multicolumn{1}{  c }{$N_n$}&
\multicolumn{1}{ c||}{$\phi_n$}\\

\hline
   3  & $1.124001... \times 10^{-2}$
      & $(1.12400 0 \pm 0.00002 1)\times 10^{   -2}$ &
                                $ 1.2 \times 10^{10}$ & 0.0580\\
   4  & $1.06838 \times 10^{-1}$ 
      & $(1.06845 4   \pm 0.00002 5  )\times 10^{   -1}$ &
                                $  2\times 10^9$ & 0.1730\\
   5  & $2.5946 \times 10^{-1}$  
      & $(2.59444     \pm 0.00007   )\times 10^{   -1}$ &
                                $  1.6\times 10^9$ & 0.3077\\
   6  & $2.9473 \times 10^{-1}$ 
      & $(2.94723 \pm 0.00009 )\times 10^{   -1}$ &
                                $  2\times 10^9$ & 0.4391\\
   7  & $1.9877 \times 10^{-1}$ 
      & $(1.98768 \pm 0.00007 )\times 10^{   -1}$ &
                                $  4\times 10^8$ & 0.5564\\
   8  & $(9.0116 \pm 0.0020)\times 10^{-2}$       
      & $(9.0131 \pm 0.0006)\times 10^{   -2}$ &
                                $  10^8$ & 0.6554\\
   9  & $(2.9644 \pm 0.0012)\times 10^{-2}$         
      & $(2.9652 \pm 0.0002)\times 10^{   -2}$ &
                                $   8\times 10^7$ & 0.7361\\
  10  & $(7.4471 \pm 0.0059)  \times 10^{-3}$         
      & $(7.4487 \pm 0.0006)\times 10^{   -3}$ &
                                $   8\times 10^7$ & 0.8002\\
  11  & $(1.4796\pm 0.0026) \times 10^{-3}$         
      & $(1.4818  \pm 0.0002)\times 10^{   -3}$ &
                                $   6\times 10^7$ & 0.8501\\
  12  & $(2.409 \pm 0.011)  \times 10^{-4}$         
      & $(2.4000  \pm 0.0002)\times 10^{   -4}$ &
                                $   6\times 10^7$ & 0.8884\\
  13  & $(3.18  \pm 0.04)   \times 10^{-5}$         
      & $(3.2324 \pm 0.0003)\times 10^{   -5}$ &
                                $   6\times 10^7$ & 0.9175\\
  14  & $(3.60  \pm 0.13)   \times 10^{-6}$         
      & $(3.6835 \pm 0.0004)\times 10^{   -6}$ &
                                $   4\times 10^7$ & 0.9393\\
  15  & $(3.7   \pm 0.4)    \times 10^{-7}$         
      & $(3.6017 \pm 0.0004)\times 10^{   -7}$ &
                                $   4\times 10^7$ & 0.9556\\
  16  & $(2.3   \pm 0.3)    \times 10^{-8}$         
      & $(3.0574 \pm 0.0004)\times 10^{   -8}$ &
                                $   4\times 10^7$ & 0.9677\\
  17  &
      & $(2.2762 \pm 0.0002)\times 10^{   -9}$ &
                                $   4\times 10^7$ & 0.9765\\
  18  & $(1.3 \pm 0.5) \times 10^{-10}$             
      & $(1.4989  \pm 0.0002 )\times 10^{  -10}$ &
                                $   4\times 10^7$ & 0.9830\\
  19  &
      & $(8.7983 \pm 0.0013)\times 10^{  -12}$ &
                                $   4\times 10^7$ & 0.9878\\
  20  & $(1.5 \pm 0.8) \times 10^{-13}$             
      & $(4.6314  \pm 0.0004 )\times 10^{  -13}$ &
                                $   8\times 10^7$ & 0.9912\\
  21  && $(2.1994 \pm 0.0004)\times 10^{  -14}$ &
                                $   2\times 10^7$ & 0.9937\\
  22  && $(9.4835 \pm 0.0017)\times 10^{  -16}$ &
                                $   2\times 10^7$ & 0.9955\\
  23  && $(3.7227 \pm 0.0005)\times 10^{  -17}$ &
                                $   2\times 10^7$ & 0.9968\\
  24  && $(1.3379 \pm 0.0003)\times 10^{  -18}$ &
                                $   2\times 10^7 $& 0.9977\\
  25  & $(9.6 \pm 5.9) \times 10^{-21}$             
      & $(4.4184 \pm 0.0004)\times 10^{  -20}$ &
                                $   4\times 10^7$ & 0.9984\\
  30  & $(1.3 \pm 1.1) \times 10^{-29}$             
      & $(5.4595  \pm 0.0005)\times 10^{  -28}$ &
                                $   4\times 10^7$ & 0.9997\\
  40  & $2.4 \times 10^{-50}$                       
      & $(6.7349 \pm 0.0006)\times 10^{  -46}$ &
                                $   8\times 10^7$ & 1.0000\\
  50  & $1.5 \times 10^{-75}$                       
      & $(5.223 \pm 0.001)\times 10^{  -66}$ &
                                $   1.6\times 10^7$ & 1.0000\\
  60  && $(7.192 \pm 0.002)\times 10^{  -88}$ &
                                $   1.2\times 10^7$ & 1.0000\\
  70  && $(3.4805 \pm 0.0004)\times 10^{ -111}$ &
                                $   3\times 10^7$ & 1.0000\\
  80  && $(9.598 \pm 0.002)\times 10^{ -136}$ &
                                $   10^7$ & 1.0000\\
  90  && $(2.1616 \pm 0.0005)\times 10^{ -161}$ &
                                $  0.8\times 10^7$ & 1.0000\\
 100  && $(5.2691 \pm 0.0006)\times 10^{ -188}$ &
                                $    1.6\times 10^7$ & 1.0000\\

 150  && $(1.0535 \pm 0.0002)\times 10^{ -332}$ &
                                $    4\times 10^6$ & 1.0000\\

 200  && $(3.818 \pm 0.001)\times 10^{ -492}$ &
                                $    4\times 10^6$ & 1.0000\\

 300  && $(1.084 \pm 0.001)\times 10^{ -841}$ &
                                $    2\times 10^6$ & 1.0000\\

 400  && $(9.863 \pm 0.003)\times 10^{-1221}$ &
                                $    4\times 10^6$ & 1.0000\\

 600  && $(3.645 \pm 0.002)\times 10^{-2040}$ &
                                $    10^6$ & 1.0000\\

 800  && $(1.326 \pm 0.001 )\times 10^{-2918}$ &
                                $    2\times 10^6$ & 1.0000\\

1000  && $(6.365  \pm 0.003)\times 10^{-3841}$ &
                                $    1.6\times 10^6$ & 1.0000\\
1200  && $(3.262  \pm 0.002)\times 10^{-4798}$ &
                                $    1.2\times 10^6$ & 1.0000\\
1400  && $(1.385  \pm 0.001)\times 10^{-5784}$ &
                                $    0.8\times 10^6$ & 1.0000\\

1600  && $(7.4306 \pm 0.0020)\times 10^{-6796}$ &
                                $    4\times 10^6$ & 1.0000\\
\hline
\end{tabular}
\caption{\footnotesize The sidedness probability $p_n$. 
Second column: literature data taken 
from Hayen and Quine \cite{HayenQuine00b} for $p_3$; 
from Calka \cite{Calka03b} for $p_4,\ldots,p_7$; 
from Brakke \cite{Brakke} for $p_8,\ldots,p_{15}$; and
from Drouffe and Itzykson \cite{DI84} for $p_n$ with $n\geq 16$.
Third column: $p_n$ and its standard deviation calculated 
by the Monte Carlo method of this work. 
Fourth column: number $N_n$ of cell generation attempts. Fifth column:
fraction $\phi_n$ of successful attempts.
}
\label{tablepn}
\end{center}

\end{footnotesize}
\end{table}

%%%%%%%%%%%%%%%%%%%%%%%%%%%%%%%%
%%%%%%%%%%%%%%%%%%%%%%%%%%%%%%%&

%%%%%%%%%%%%%%%%%%%%%%%%%%%%%%%%
%%%%%%%%%%%%%%%%%%%%%%%%%%%%%%%&
\begin{table}
\begin{footnotesize}

\begin{center}
\renewcommand{\arraystretch}{1.1}
\begin{tabular}{|| r | r r r r l ||}
\hline

\multicolumn{1}{||r| }{$n$}&
\multicolumn{2}{ |c  }{$\la P \ra_n$}&
\multicolumn{2}{  c  }{$\la P^2 \ra_n$}&
\multicolumn{1}{  c| }{$\la \delta P^2 \ra_n^{1/2} $}
\\
\hline
   3 &    2.740296 &(2) &    8.17130 &(2) &    0.81368
          \\
   4 &    3.219524 &(3) &   11.04819 &(2) &    0.82634
          \\
   5 &    3.642658 &(3) &   13.96626 &(3) &    0.83504
          \\
   6 &    4.026307 &(4) &   16.91958 &(4) &    0.84169
          \\
   7 &    4.380000 &(6) &   19.90272 &(6) &    0.84702
          \\
   8 &    4.710196 &(8) &   22.91084 &(8) &    0.85140
          \\
   9 &    5.020869 &(12) &   25.94026 &(12) &    0.85506
          \\
  10 &    5.315211 &(11) &   28.98790 &(12) &    0.85816
          \\
  11 &    5.595488 &(10) &   32.05043 &(12) &    0.86078
          \\
  12 &    5.863536 &(11) &   35.12588 &(13) &    0.86304
          \\
  13 &    6.12072 &(2) &   38.2114 &(2) &    0.86497
          \\
  14 &    6.36824 &(1) &   41.3055 &(2) &    0.86664
          \\
  15 &    6.60705 &(2) &   44.4066 &(2) &    0.86809
          \\
  16 &    6.83797 &(2) &   47.5136 &(3) &    0.86936
          \\
  17 &    7.06173 &(2) &   50.6258 &(2) &    0.87047
          \\
  18 &    7.27884 &(2) &   53.7410 &(3) &    0.87145
          \\
  19 &    7.48992 &(2) &   56.8598 &(3) &    0.87233
          \\
  20 &    7.69544 &(1) &   59.9820 &(3) &    0.87310
          \\
  21 &    7.89576 &(3) &   63.1066 &(4) &    0.87380
          \\
  22 &    8.09118 &(3) &   66.2318 &(5) &    0.87442
          \\
  23 &    8.28215 &(2) &   69.3596 &(3) &    0.87499
          \\
  24 &    8.46892 &(3) &   72.4890 &(4) &    0.87551
          \\
  25 &    8.65171 &(2) &   75.6198 &(4) &    0.87598
          \\
  30 &    9.51379 &(2) &   91.2825 &(5) &    0.87783
          \\
  40 &   11.03971 &(1) &  122.6501 &(4) &    0.88005
          \\
  50 &   12.37983 &(2) &  154.0370 &(5) &    0.88135
          \\
  60 &   13.58887 &(3) &  185.4355 &(7) &    0.88219
          \\
  70 &   14.69896 &(2) &  216.8384 &(6) &    0.88279
          \\
  80 &   15.73105 &(2) &  248.2460 &(7) &    0.88323
          \\
  90 &   16.69951 &(3) &  279.6545 &(8) &    0.88357
          \\
 100 &   17.61487 &(2) &  311.0645 &(8) &    0.88384
          \\
 150 &   21.61817 &(3) &  468.1280 &(12) &    0.88465
          \\
 200 &   24.98833 &(3) &  625.1998 &(11) &    0.88505
          \\
 300 &   30.63607 &(2) &  939.3527 &(12) &    0.88544
          \\
 400 &   35.39384 &(2) & 1253.5092 &(13) &    0.88564
          \\
 600 &   43.37090 &(4) & 1881.820 &(3) &    0.88584
          \\
 800 &   50.09346 &(2) & 2510.143 &(2) &    0.88593
          \\
1000 &   56.01490 &(1) & 3138.456 &(2) &    0.88599
          \\
1200 &   61.36764 &(1) & 3766.773 &(2) &    0.88603
          \\
1400 &   66.28953 &(2) & 4395.087 &(4) &    0.88606
          \\
1600 &   70.87047 &(2) & 5023.408 &(2) &    0.88608
         \\
$\infty$&           &    &          &    &    0.886226...
         \\
\hline
\end{tabular}
\caption{\footnotesize Estimates of the average $\la P \ra_n$,
the second moment $\la P^2 \ra_n$, and the root-mean-square fluctuation
$\la \delta P^2 \ra_n^{1/2}$ of the cell perimeter $P$.
The numbers in parentheses represent the standard deviation in the last digit.
The entries of the third column have an error of at most one unit in their
last digit. The limit value $\tfrac{1}{2}\pi^{1/2}=0.886226...$ for $n=\infty$
has the status of a conjecture.
}
\label{tableP}
\end{center}

\end{footnotesize}
\end{table}
%%%%%%%%%%%%%%%%%%%%%%%%%%%%%%%%
%%%%%%%%%%%%%%%%%%%%%%%%%%%%%%%&

%%%%%%%%%%%%%%%%%%%%%%%%%%%%%%%%
%%%%%%%%%%%%%%%%%%%%%%%%%%%%%%%&
\begin{table}
\begin{footnotesize}

\begin{center}
\renewcommand{\arraystretch}{1.1}
\begin{tabular}{|| r | r r r r l ||}
\hline
\multicolumn{1}{||r| }{$n$}&
\multicolumn{2}{ |c  }{$\la A \ra_n$}&
\multicolumn{2}{  c  }{$\la A^2 \ra_n$}&
\multicolumn{1}{  c||}{$n^{-1/2}\la \delta A^2 \ra_n^{1/2} $}
\\
\hline
   3 &    0.343087 &           (3) &    0.161573 &   (3) 
          & 0.12092\\
   4 &    0.558052 &           (4) &    0.401285 & (5) 
          & 0.14989\\
   5 &    0.774080 &           (4) &    0.73675 &           (1) 
          & 0.16586\\
   6 &    0.995789 &            (5) &    1.17953 &       (2) 
          & 0.17698\\
   7 &    1.22251 &  (1) &    1.73516 & (3) 
          & 0.18541\\
   8 &    1.45328 & (1) &    2.40724 &(3) 
          & 0.19200\\
   9 &    1.68736 & (1) &    3.19847 &(4) 
          & 0.19756\\
  10 &    1.92408 &(2) &    4.11064 &(7) 
          & 0.20213\\
  11 &    2.16295 &(2) &    5.1451 &(1)  
          & 0.20599\\
  12 &    2.40366 &(2) &    6.3033 &(1)  
          & 0.20929\\
  13 &    2.64578 &(2) &    7.5854 &(1) 
          & 0.21217\\
  14 &    2.88906 &(3) &    8.9920 &(1)  
          & 0.21469\\
  15 &    3.13331 &(3) &   10.5234 &(1)  
          & 0.21689\\
  16 &    3.37835 &(3) &   12.1797 &(2)  
          & 0.21885\\
  17 &    3.62416 &(2) &   13.9619 &(2)  
          & 0.22059\\
  18 &    3.87034 &(3) &   15.8680 &(2)  
          & 0.22215\\
  19 &    4.11703 &(3) &   17.8996 &(3)  
          & 0.22357\\                  
  20 &    4.36415 &(3) &    20.0570 &(3) 
          & 0.22484\\
  21 &    4.61158 &(4) &   22.3394 &(3)  
          & 0.22601\\
  22 &    4.85923 &(4) &   24.7464 &(4)  
          & 0.22706\\
  23 &    5.10715 &(4) &   27.2790 &(4)  
          & 0.22803\\
  24 &    5.35531 &(5) &   29.9371 &(5)  
          & 0.22891\\
  25 &    5.60358 &(6) &   32.7198 &(6) 
          & 0.22974\\
  30 &    6.84686 &(6) &    48.5090 &(6) 
          & 0.23306\\
  40 &    9.33913 &(4) &    89.470 &(1) 
          & 0.23723\\
  50 &   11.83458 &(7) &   142.932 &(2) 
          & 0.23977\\
  60 &   14.33183 &(6) &   208.900 &(2) 
          & 0.24146\\
  70 &   16.82979 &(6) &   287.363 &(3) 
          & 0.24267\\
  80 &   19.3283 &(1) &   378.331 &(4) 
          & 0.24358\\
  90 &   21.82726 &(8) &   481.800 &(4) 
          & 0.24429\\
 100 &   24.32627 &(6) &   597.763 &(3) 
          & 0.24485\\
 150 &   36.8236 &(2) &  1365.10 &(1)   
          & 0.24658\\
 200 &   49.3224 &(1) &  2444.94 &(1)   
          & 0.24742\\
 300 &   74.3214 &(2) &  5542.16 &(3)   
          & 0.24826\\
 400 &   99.3206 &(1) &  9889.33 &(3) 
          & 0.24871\\
 600 &  149.3196 &(3) & 22333.58 &(1) 
          & 0.24914\\
 800 &  199.3201 &2) & 39778.23 &(6) 
          & 0.24935\\
1000 &  249.3192 &(2) & 62222.3 &(1) 
          & 0.24948\\
1200 &  299.3193 &(2) & 89666.7 & (1) 
          & 0.24957\\
1400 &  349.3187 &(3) & 122110.8 & (2) 
          & 0.24963\\
1600 &  399.3190 &(2) &159555.4 & (2) 
          & 0.24968\\
$\infty$ &       &    &         &
          & 0.250000...\\
\hline
\end{tabular}
\caption{\footnotesize  Estimates of the average $\la A \ra_n$,
the second moment $\la A^2 \ra_n$, and the normalized
root-mean-square fluctuation
$n^{-1/2} \la \delta A^2 \ra_n^{1/2}$ of the cell area $A$.
The numbers in parentheses represent the standard deviation in the last digit.
The entries of the third column have an error of at most one unit in their
last digit. The limit value $\tfrac{1}{4}$ for $n=\infty$ 
has the status of a conjecture. 
}
\label{tableA}
\end{center}

\end{footnotesize}
\end{table}
%%%%%%%%%%%%%%%%%%%%%%%%%%%%%%%%
%%%%%%%%%%%%%%%%%%%%%%%%%%%%%%%&

%%%%%%%%%%%%%%%%%%%%%%%%%%%%%%%%%%%%%%%%%%%%%%%%%%%%%%%%%%%%%%%%%%%%%%%%%%%%

\subsection{Perimeter and area}
\label{secresultsav}

We have simulated the two cell properties
that have received the greatest attention in the literature, {\it viz.}
the cell perimeter $P$ and the cell area $A$. 
We determined the average, the
second moment and the variance of both of these quantities as a function of
$n$. The perimeter results are summarized in Table 
\ref{tableP} 
and the area results in Table 
\ref{tableA}. 

Similar tables extracted from the literature 
were compiled by Okabe {\it et al.} \cite{Okabeetal00}.
However, by far the most accurate ones 
appear in unpublished work by Brakke \cite{Brakke} 
and concern the regime $3 \leq n \leq 16$. 
All our area and perimeter data 
are compatible with those of Ref.\,\cite{Brakke},
but our error bars are strongly reduced.
A further check on the numerical data is provided by two more sum rules,
\beq
\overline{P_n}=4.000\,035 (65), \qquad \overline{A_n}=1.000\,02 (2), 
\label{sumrulesbis}
\eeq
for which the exact values are $4$ and $1$, respectively.

%%%%%%%%%%%%%%%%%%%%%%%%%%%%%%%%
%%%%%%%%%%%%%%%%%%%%%%%%%%%%%%%%
%%%%%%%%%%%%%%%%%%%%%%%%%%%%%%%%
%%%%%%%%%%%%%%%%%%%%%%%%%%%%%%%%

%%%%%%%%%%%%%%%%%%%%%%%%%%%%%%%%
%%%%%%%%%%%%%%%%%%%%%%%%%%%%%%%%
\begin{figure}
\begin{center}
\scalebox{.60}
{\includegraphics{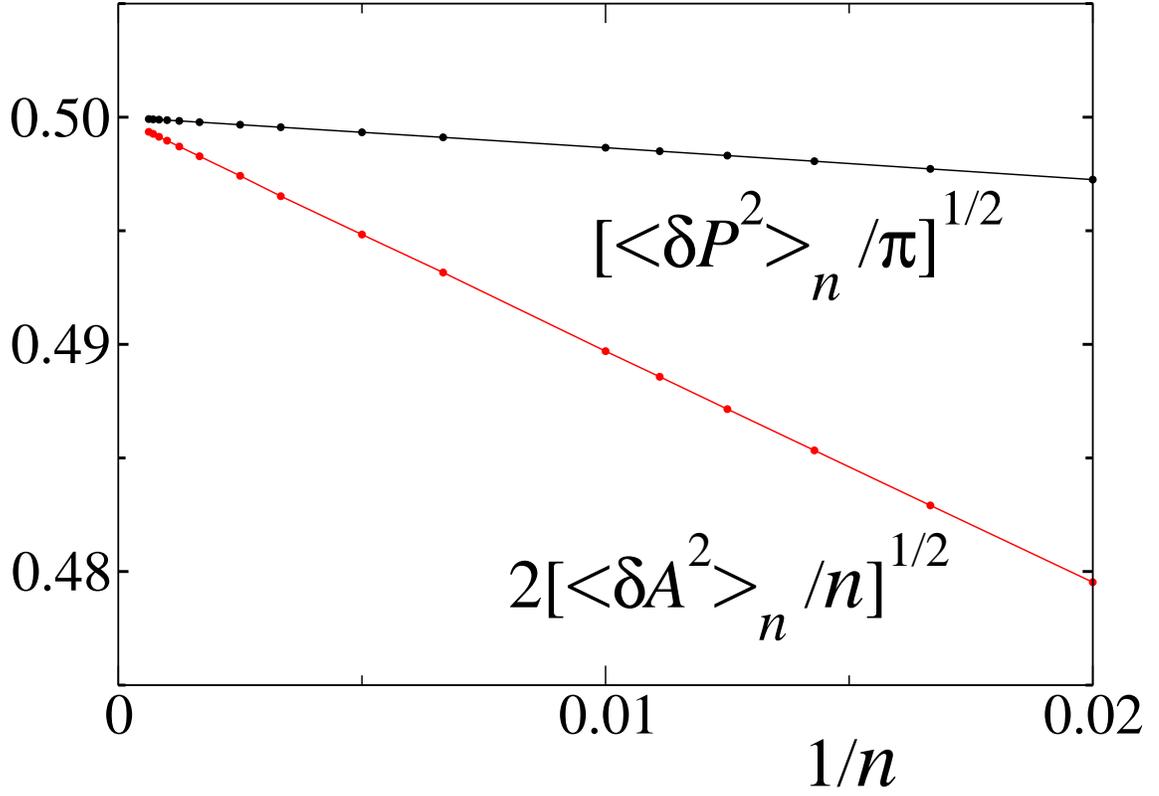}}
\end{center}
\caption{{\small The asymptotic large-$n$ behavior of
the root mean square fluctuations $\la \delta P^2 \ra_n^{1/2}$
and $\la \delta A^2 \ra_n^{1/2}$
of the perimeter and the area, respectively, of the $n$ sided cell.
Data shown are in the range $50 \leq n \leq 1600$.
The error bars are smaller than the data points.
The two curves that connect the data points are asymptotically
straight lines which for $n\to\infty$ both 
appear to converge $\half$.
}}
\label{figvare}
\end{figure}
%%%%%%%%%%%%%%%%%%%%%%%%%%%%%%%%
%%%%%%%%%%%%%%%%%%%%%%%%%%%%%%%%

We now turn to the large $n$ behavior.
Our data indicate the expansions
\bea
\la P \ra_n   &=& (\pi n)^{\half} + a_{\half}\,n^{-\half} +
 a_{\frac{3}{2}}\,n^{-\frac{3}{2}} + \ldots, \nonumber\\
\la P^2 \ra_n &=& \pi n + b_0 + b_1n^{-1} + \ldots, \nonumber\\
\la A \ra_n   &=&  \tfrac{1}{4}n + c_0 + c_1n^{-1} + \ldots, \nonumber\\
\la A^2 \ra_n &=& (\tfrac{1}{4}n)^2 + d_{-1}n + d_0 + \ldots\,,
\label{expansions}
\eea
which again go down by integer powers of $n$. 
They imply that the variances have the series
\bea
\la P^2 \ra_n - \la P \ra_n^2 &=& b_0 - 2\pi^{\half}a_{\half} + \ldots
\nonumber\\
\la A^2 \ra_n - \la A \ra_n^2 &=& (d_{-1} -\tfrac{1}{2}c_0) n + \ldots\,.
\label{expvariances}
\eea
The leading terms in each of the four series of Eq.\,(\ref{expansions}) 
are known from theoretical analysis \cite{HJHletter05,HJHart05}.
Heuristically they follow from the sole observation that in the large $n$
limit the Voronoi cell becomes a circle of radius $R_{\rm c}=(n/4\pi)^{1/2}$. 
Theoretical analysis
can in principle also produce the higher order terms in (\ref{expansions}), 
but this has not been attempted yet.
Consequently, the leading coefficients of the series in (\ref{expvariances})
are not known analytically. Here our simulations results provide answers.

In Fig.\,\ref{figvare} we have plotted $[\la \delta P^2 \ra_n/\pi]^{1/2}$
and $2[\la \delta A^2 \ra_n/n]^{1/2}$.
The numerical data strongly point to limit values
equal to $\frac{1}{2}$ for both quantities when $n\to\infty$.
Conjecturing that these limits are exact we then conclude that
\beq
b_0-2\pi^{\half}a_{\half} = \tfrac{1}{4}\pi, \qquad
d_{-1}-\tfrac{1}{2}c_0  = \tfrac{1}{16}.
\label{conjecture}
\eeq
Analysis of the $\la P^2 \ra_n$ data to next order suggests that
$\la P^2 \ra_n/(\pi n) - 1$ tends to $-1$ as $n\to\infty$.
Conjecturing that this, too, is exact and combining it with the first one of
Eqs.\,(\ref{conjecture}) we arrive at
\beq
a_{\half} = -\tfrac{5}{8}\pi^{\half}, \qquad b_0=-\pi,
\label{abanal}
\eeq
We do not attempt, however, a similar analytical conjecture for the
second pair of coefficients, $c_0$ and $d_{-1}$,
nor will we pursue estimates for the higher order coefficients in the series 
(\ref{expansions}) and (\ref{expvariances}),
except below in connection with Lewis' law.
\vspace{3mm}

Lewis' celebrated law \cite{Lewis} is an empirical
statement about one of the cell's 
most conspicuous properties, {\it viz.}~the relation between 
its area and its number of sides.
The law states that
the average area $\la A\ra_n$ of an $n$-sided cell 
increases with $n$ as 
\beq
\la A\ra_n=\frac{a_0}{\lambda}(n-n_0),
\label{eqnAn}
\eeq
where $a_0$ and $n_0$ are constants and where we have displayed again the
dependence on the areal particle density $\lambda$. 
Sometimes (see the discussion in Ref.\,\cite{Okabeetal00})
this law is written in the more restricted one-parameter form
\beq
\la A \ra_n = \frac{b(n-6)-1}{\lambda}
\label{eqnAnbis}
\eeq
It is found, however, that $\la A\ra_n$ deviates from linearity with $n$
in simulations of Poisson-Voronoi diagrams
as well as in the data from most experimental systems. 
We now look at what the asymptotic analysis has to say. 

In Refs.\,\cite{HJHletter05,HJHart05}
we proved that asymptotically
\beq
\la A\ra_n \simeq \pi R_c^2 
= \frac{n}{4\lambda}, \qquad n\to\infty.
\label{Anaspt}
\eeq
and this result has been incorporated in the series for $\la A\ra_n$ in
(\ref{expansions}). 
A coefficient $a_0 \approx \frac{1}{4}$
had since long been suspected by various authors \cite{DI84,MM82,LMH93}.
Going now beyond (\ref{Anaspt}) and proceeding in the same way as for $p_n$, 
we can determine
the coefficients of the series of (\ref{expansions}) for $\la A \ra_n$ 
on the basis of our simulation results of Table \ref{tableP}. 
This yields
\beq
c_0=-0.6815(5), \quad c_1=0.750(5), \quad c_2=3.15(10).
\label{valuesci}
\eeq
We now consider the laws (\ref{eqnAn}) and (\ref{eqnAnbis}).
The fact that we found $c_1$, $c_2,\ldots$\, to be nonvanishing
confirms once more that $\la A \ra_n$ is not strictly linear in $n$.
From the above it follows that in (\ref{eqnAn}) one should choose
\beq
n_0 = -4c_0 = 2.7260 (4)
\label{n0num}
\eeq
if one wishes it to correctly represent the asymptotic 
behavior of $\la A\ra_n$ for Poisson-Voronoi cells.
This is of course different from finding a best fit to a limited set
of $\la A\ra_n$ data in a restricted $n$ interval. 
If that is the purpose,
the values of $a_0$ and $n_0$ will depend on the available data
and on the way the fit is carried out.
The one-parameter law (\ref{eqnAnbis}) 
postulates a relation between $a_0$ and $n_0$ that is violated
in the asymptotic expansion. Hence (\ref{eqnAnbis}) cannot be used to describe
the large $n$ behavior of $\la A\ra_n$ and merely has the status of
an empirical fit to the data, the value of $b$ again depending on the data
set and on how the fit is done.

%%%%%%%%%%%%%%%%%%%%%%%%%%%%%%%%%%%%%%%%%%%%%%%%%%%%%%%%%%%%%%%%%%%%%%%%%%%%%

\section{Characteristic cell shapes}
\label{secshapes}

It has been established \cite{HJHletter05,HJHart05}
that in the large $n$ limit the
Voronoi cell tends to a disk of radius $R_{\rm c}=(n/4\pi)^{1/2}$
\cite{footnotea}.
In Ref.\,\cite{HJHart05} it was furthermore shown that the cell perimeter 
undergoes `elastic' deformations from circularity, 
the elasticity being, of course, of entropic origin.
The probability law of these deformations was given in the large $n$ limit.
In this section we show how our Monte Carlo method 
allows us what was hitherto impossible, 
namely to simulate for any finite $n$
the detailed statistics of the cell shape.

We Monte Carlo generated cells of prescribed sidedness $n$
in a `natural', that is, an unbiased, environment.
This was done as follows.
For a given $n$ the cell angles $(\xi,\eta)$ were drawn randomly and
$\beta_*$ was found according to the
rules of section \ref{secpn}.
The cell radius was taken equal to its most probable value 
$R_{\rm c}=(n/4\pi)^{1/2}$
and the cell boundary was constructed. 
This boundary, together with the position of the central particle,
fixes the positions of the $n$ first neighbor particles.
We then determined the cell's fundamental domain ${\cal F}$, that is,
the union of the $n$ disks of radius $S_m$ centered at the vertices $\bS_m$.
The complement of ${\cal F}$ in a large rectangle of suitable size
was subsequently filled randomly with particles 
of a uniform density equal to one. The particles added by this procedure
are all necessarily second or higher order neighbors of the central one.
The Voronoi construction was finally applied to the full collection
of particles to complete the Voronoi cell diagram.

%%%%%%%%%%%%%%%%%%%%%%%%%%%%%%%%%%%%%%%%%%%%%%%%%%%%%%%%%%%%%%%%%%%%%%%%%%%%%%

\subsection{Cells of $n=24,\, 48,$ and $96$ sides}
\label{secshapes1}

\begin{figure}
\begin{center}
\scalebox{.60}
{\includegraphics{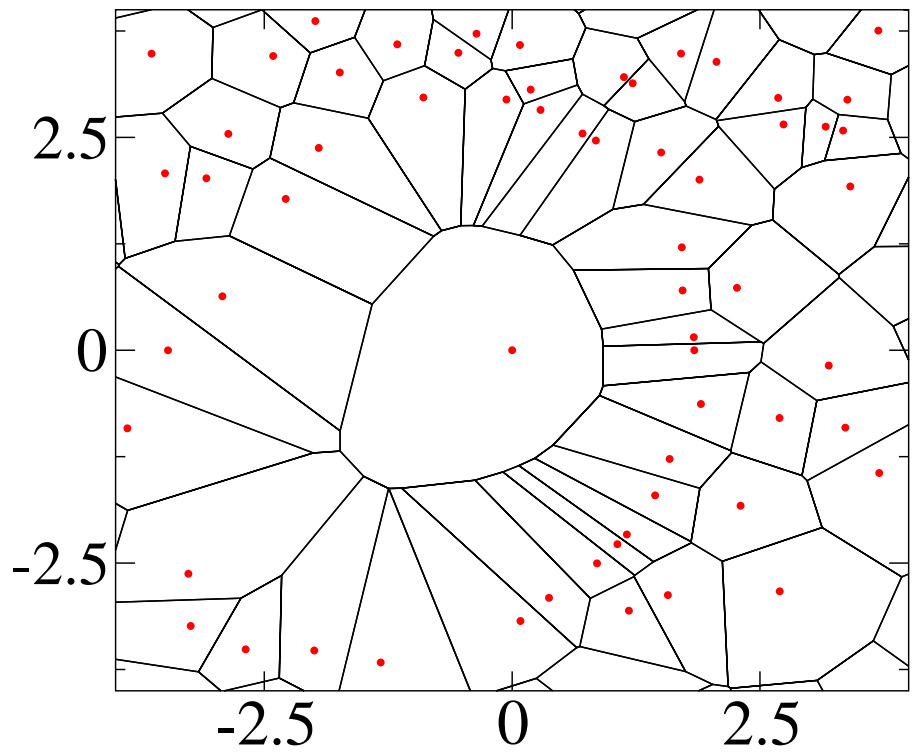}}
\end{center}
\caption{\small A typical Voronoi cell with $n=24$ neighbors.
The dots represent the particles.}
\label{fig24}
\end{figure}

\begin{figure}
\begin{center}
\scalebox{.60}
{\includegraphics{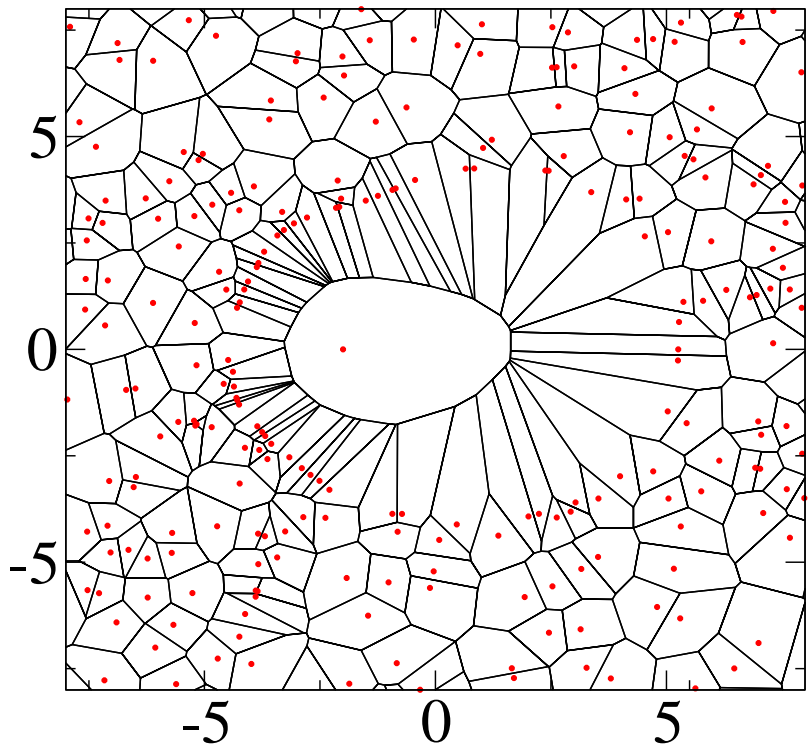}}
\end{center}
\caption{\small A typical Voronoi cell with $n=48$ neighbors.}
\label{fig48}
\end{figure}

\begin{figure}
\begin{center}
\scalebox{.60}
{\includegraphics{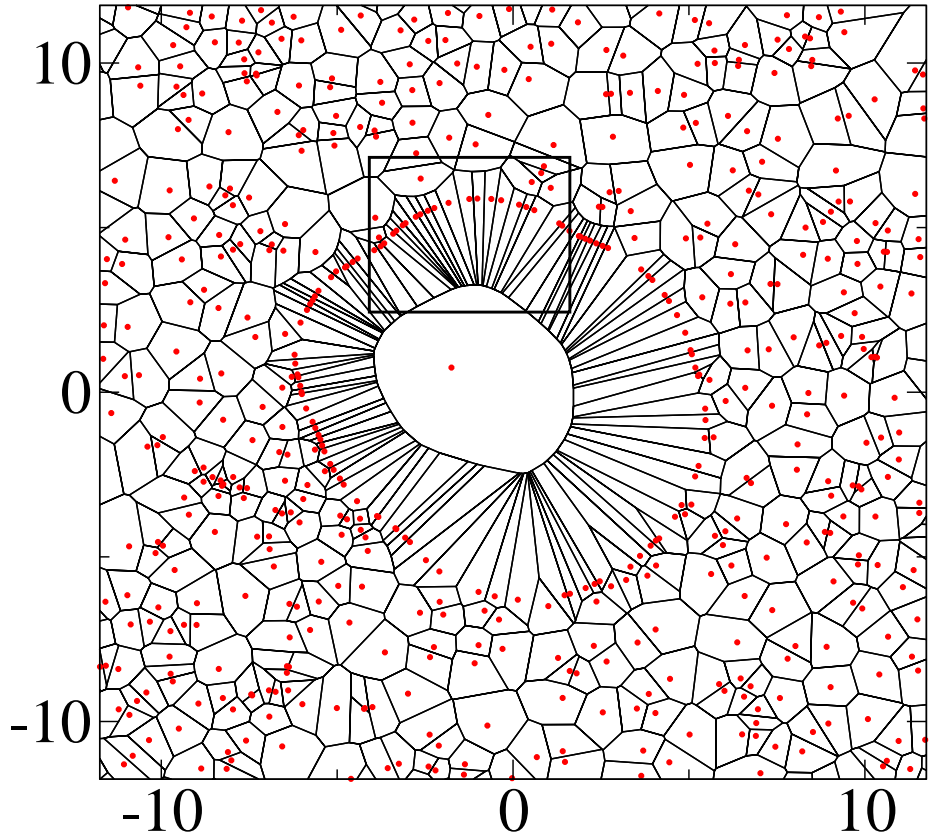}}
\end{center}
\caption{\small A typical Voronoi cell with $n=96$ neighbors.
The region inside the box is shown enlarged in Fig.\,{\ref{fig96detail}}.}
\label{fig96}
\end{figure}

We have generated typical cell shapes for a sequence of values of $n$,
starting with $n=3$ and doubling $n$ each time.
Figs.\,\ref{fig24}, \ref{fig48}, and $\ref{fig96}$,
in which the dots represent the particles,
show the results for cells of 
$n=24$, $48$, and $96$ neighbors.
The three pictures are at different scales but all have 
unit particle density.
This picture sequence illustrates
the tendencies that characterize many-sided cells.
One tendency is for the first neighbor cells to be
elongated. This feature is apparent already for $n=24$
and becomes very pronounced for $n=48$, whereas the $n=96$ cell has
{\it only} very elongated neighbors.
The same phenomenon was observed by Lauritsen {\it et al.} \cite{LMH93},
but in a different system.
These authors consider Poisson-Voronoi diagrams 
to which they assign an `energy' that favors
many-sided cells. Snapshots of 
their configurations show a dense structure of many-sided cells
(of sidednesses higher than $n=60$) 
separated by mostly four-sided elongated cells.
Their procedure 
does not, however, provide estimates for $p_n$ in an unbiased 
Poisson-Voronoi diagram.

Another tendency, appearing similarly in Ref.\,\cite{LMH93}, 
is for the first-neighbor particles to align on what
tends towards a continuous curve. Whereas for $n=24$ some imagination is still
needed to see this curve, it
becomes clearly distinguishable for $n=48$ and is immediately obvious to the
eye for $n=96$.  The typical distance between nearest neighbor particles
along this curve decreases as $2\pi(2R_{\rm c})/n \sim n^{-1/2}$. 
We note that whereas Voronoi cells are always convex,
the `curve' connecting the first neighbors need not enclose a convex area;
in each of the Figs.\,\ref{fig24}, \ref{fig48}, and \ref{fig96} 
there are small but clearly
distinguishable deviations from convexity. 

Fig.\,\ref{fig96detail} enlarges a detail of Fig.\,\ref{fig96}
and shows a collection of first-neighbor cells. All first neighbors
fully visible in the
figure are four-sided except those marked $A$, $B$, $D$, $E$, which are
five-sided, and the cell $C$, which is either five- or six-sided
(this depends on how the two almost coinciding three-vertices are arranged
at the point marked `2V'; a higher resolution is needed to decide this
question). Fig.\,\ref{fig96detail} illustrates 
that in the limit of large $n$
four-sided first neighbors become dominant. 
In Ref.\,\cite{HJHaboav06} it was argued that five-sided cells 
constitute a fraction only of order $n^{-1/2}$ of
all first-neighbor cells, and that the probability of
six- and higher-sided first-neighbors is of still higher order in $n^{-1/2}$. 
In Fig.\,\ref{fig96detail} the cell marked $P$ is a second neighbor to the
central cell. The boundary separating it from the first neighbors
has been shown as a heavy solid line on which we will further comment shortly.

%%%%%%%%%%%%%%%%%%%%%%%%%%%%%%%%%%%%%%%%%%%%%%%%%%%%%%%%%%%%%%%%%%%%%%%%%%%%%

\subsection{Very large cells}
\label{secsides2}

Focusing now on
very large $n$ we show in Fig.\,\ref{fig1536} 
a central particle located in the origin and
having $1536$ neighbors. As before, the dots represent the particles. 
The inner contour, which is
nearly indistinguishable from a circle of radius $R_{\rm c}$, 
represents the boundary of the Voronoi cell of the central particle.
The outer `curve', which is also very close to circular but of radius 
$2R_{\rm c}$, represents the
alignment of the $1536$ first-neigbor particles.
Their high line density gives the impression of a continuous curve.
Cell boundaries other than those of the central cell
have not been drawn; they would totally blacken the empty
annular region between the boundary of the central cell 
and its first-neighbor particles.

The boxed region
in Fig.\,\ref{fig1536} is shown enlarged in
Fig.\,\ref{fig1536detail}, where we did draw all Voronoi cell boundaries. 
The extreme elongation of the first-neighbor cells is what first 
strikes the eye. 
The discrete structure of the `curve' of first neighbors is 
also apparent now.
The distances $\ell_m$ between successive first-neighbor particles along this
curve are of order $n^{-1/2}$.
More precisely, if we set $\ell_m=\lambda_m(4\pi/n)^{1/2}$,
then the theory \cite{HJHart05} implies that for $n\to\infty$ the $\lambda_m$
are independent identically distributed random variables of probability law
$\lambda_m\exp(-\lambda_m)$.
Random deviations from a local straight line 
are too small to be discernible to the eye;
they may be argued \cite{HJHunpublished} to decrease as $n^{-3/2}$,
which is also the order of magnitude of the systematic deviations due to the
radius of curvature $2R_{\rm c}$. 
The large cell and its environment are characterized, therefore, by
four different length scales, each varying with its own power of $n$. They
have been summarized in Table \ref{tablescales} below. 
One has to go to $n$ values as high as
we did in order for the separation of scales to become clearly visible.

\begin{table}
\begin{footnotesize}

\begin{center}
\renewcommand{\arraystretch}{1.1}
\begin{tabular}{|| l  l ||}
\hline
\multicolumn{1}{|| c  }{Scale}&
\multicolumn{1}{   c||}{Length}\\
\hline
$n^{1/2}$  &  Cell radius\\

$n^0$    &  Typical interparticle distance outside the first neighbor circle\\

$n^{-1/2}$ &  Typical distance between successive first neighbor particles\\

$n^{-3/2}$ &  Random deviations of first neighbors from full alignment\\
\hline
\end{tabular}
\caption{\footnotesize  Four length scales characterizing the $n$-sided
Voronoi cell in the large $n$ limit; see Figs.\,\ref{fig1536} and
\ref{fig1536detail}.
}
\label{tablescales}
\end{center}

\end{footnotesize}
\end{table}

\begin{figure}
\begin{center}
\scalebox{.60}
{\includegraphics{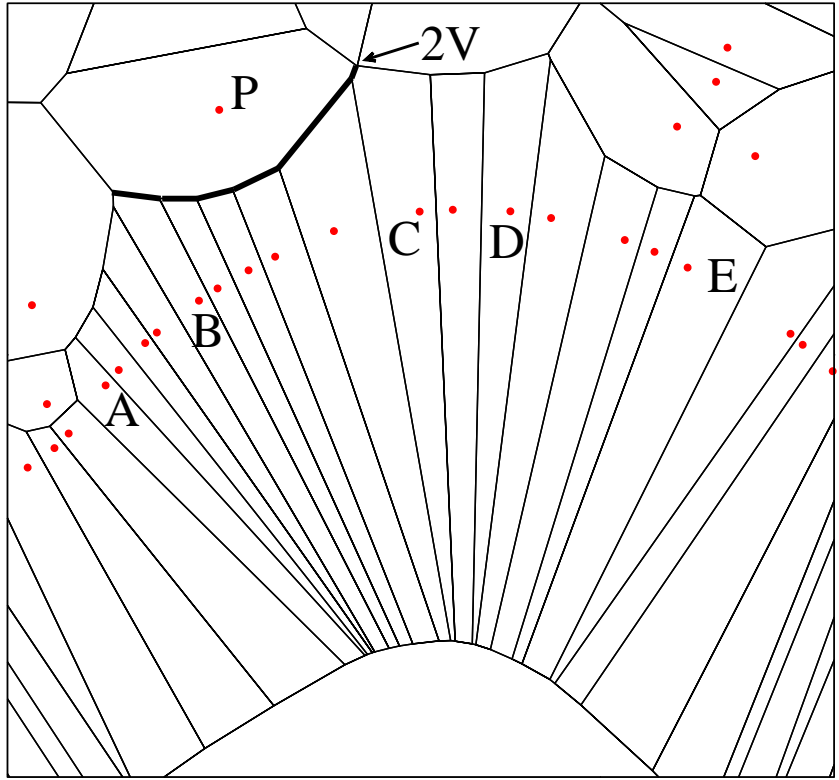}}
\end{center}
\caption{\small 
Enlargement 
of the box in Fig.\,\ref{fig96} showing some of the strongly
elongated first neighbors of the central cell.
Among the first neigbors fully visible, cells $A, B, C, D,$ and $ E$ 
have more than four sides. 
The arrow marked `2V' points to 
two three-vertices that coincide within the resolution
of the figure.
Cell $P$ is a second neighbor whose boundary with the first neighbors
(heavy lines) is an example of an `incipient parabola segment' (see text). 
}
\label{fig96detail}
\end{figure}

\begin{figure}
\begin{center}
\scalebox{.60}
{\includegraphics{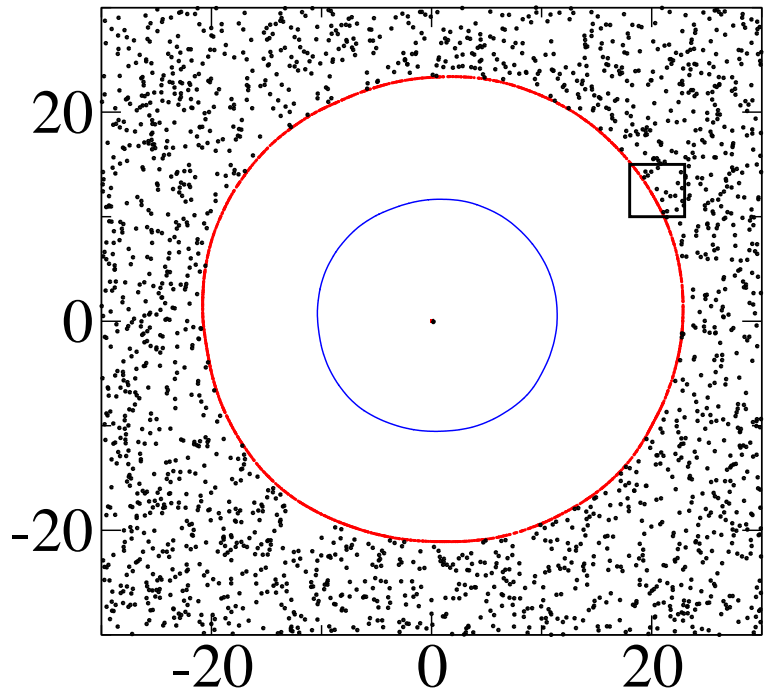}}
\end{center}
\caption{\small 
Approach to the infinite $n$ limit.
The origin is occupied by a particle whose Voronoi cell has $n=1536$ sides.
The almost circular inner curve is the cell boundary of the central cell.
The other cells boundaries have not been drawn. 
The almost circular outer curve is made up of the $1536$ first-neighbor
particles. 
The region inside the box is shown enlarged in Fig.\,\ref{fig1536detail}.
}
\label{fig1536}
\end{figure}

\begin{figure}
\begin{center}
\scalebox{.60}
{\includegraphics{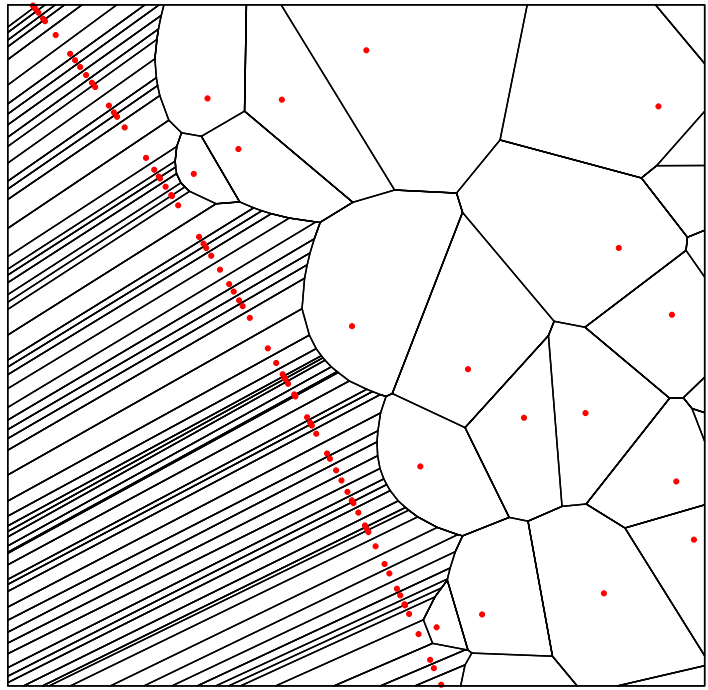}}
\end{center}
\caption{\small 
Enlargement of the box in Fig.\,\ref{fig1536}, where now all cell boundaries
have been drawn. The discrete structure of the outer `curve' of
Fig.\,\ref{fig1536} has become visible here.
}
\label{fig1536detail}
\end{figure}

Very large $n$ is required also for
still another feature to become apparent.
In Fig.\,\ref{fig1536detail} 
the boundary between a second-neighbor cell and its adjacent
first-neighbor cells is, by construction, 
composed of points equidistant to the second-neighbor particle
and the almost continuous straight
line of first-\-neighbor particles. But such a boundary is a parabola.
Hence, in the limit $n\to\infty$ the boundary separating 
the set of first from the set of second-neigbor cells 
is {\it piecewise parabolic}.
Indeed, with this observation in mind one now recognizes
the boundary segment of cell $P$ in Fig.\,\ref{fig96detail}
(heavy solid line), and others in that same figure, as `incipient parabolic'. 
Such knowledge was at the basis of the theory
of two-cell correlations exposed in Ref.\,\cite{HJHaboav06}.
There, laws discovered in the $n\to\infty$ limit were
extrapolated backward and shown to be relevant for finite $n$.
It was shown, in particular, that Aboav's linear relationship
\cite{Aboav70,Okabeetal00} between $n$ and
the total average sidedness $nm_n$ of the neighbors of an $n$-sided cell
cannot hold in Poisson-Voronoi diagrams.
We expect that in the future the study of large cells will shed 
further light also on various
issues relevant for the finite $n$ behavior. 

%%%%%%%%%%%%%%%%%%%%%%%%%%%%%%%%
%%%%%%%%%%%%%%%%%%%%%%%%%%%%%%%%
%%%%%%%%%%%%%%%%%%%%%%%%%%%%%%%%
%%%%%%%%%%%%%%%%%%%%%%%%%%%%%%%%

%%%%%%%%%%%%%%%%%%%%%%%%%%%%%%%%
%%%%%%%%%%%%%%%%%%%%%%%%%%%%%%%%
%%%%%%%%%%%%%%%%%%%%%%%%%%%%%%%%
%%%%%%%%%%%%%%%%%%%%%%%%%%%%%%%%

%%%%%%%%%%%%%%%%%%%%%%%%%%%%%%%%
%%%%%%%%%%%%%%%%%%%%%%%%%%%%%%%%
%%%%%%%%%%%%%%%%%%%%%%%%%%%%%%%%
%%%%%%%%%%%%%%%%%%%%%%%%%%%%%%%%

%%%%%%%%%%%%%%%%%%%%%%%%%%%%%%%%
%%%%%%%%%%%%%%%%%%%%%%%%%%%%%%%%
%%%%%%%%%%%%%%%%%%%%%%%%%%%%%%%%
%%%%%%%%%%%%%%%%%%%%%%%%%%%%%%%%

%%%%%%%%%%%%%%%%%%%%%%%%%%%%%%%%
%%%%%%%%%%%%%%%%%%%%%%%%%%%%%%%%
%%%%%%%%%%%%%%%%%%%%%%%%%%%%%%%%
%%%%%%%%%%%%%%%%%%%%%%%%%%%%%%%%

%%%%%%%%%%%%%%%%%%%%%%%%%%%%%%%%
%%%%%%%%%%%%%%%%%%%%%%%%%%%%%%%%
%%%%%%%%%%%%%%%%%%%%%%%%%%%%%%%%
%%%%%%%%%%%%%%%%%%%%%%%%%%%%%%%%

%%%%%%%%%%%%%%%%%%%%%%%%%%%%%%%%%%%%%%%%%%%%%%%%%%%%%%%%%%%%%%%%%%%%%%%%%%%%%%

\section{Summary and conclusion}
\label{secconclusion}

In this paper we have developed 
a new Monte Carlo method for evaluating
the sidedness probability $p_n$ for arbitrary $n$.
The method, which is constructed on the basis of an extension of earlier
theory \cite{HJHletter05,HJHart05}, is not difficult to implement
once one has available the rather complicated
analytic expressions that intervene.

We have determined $p_n$ as well as 
the first and second moments of the $n$ dependent
cell perimeter and cell area.
Full agreement is obtained with earlier results for $p_n$ due to 
Hayen and Quine \cite{HayenQuine00b}, Calka \cite{Calka03b}, 
and Brakke \cite{Brakke}, whose data extend up to $n=16$.
For $n \gsim 10$ we have reduced the error bars on $p_n$ very considerably.
In the range up to $n=50$ we improved and corrected the $p_n$ data 
due to Drouffe and Itzykson \cite{DI84}.
For $50 < n \leq 1600$ we obtained data in a range that
that had so far remained inaccessible.
This enabled us to
investigated the asymptotic large $n$ behavior 
of $p_n$ and of the perimeter and area moments. 
On the basis of our numerical results we conclude that
they all have asymptotic series in entire powers of $n^{-1}$,
possibly up to an overall prefactor $n^{\half}$. 

Exploiting our full control of the cell statistics 
we have exhibited occurrences of extremely rare many-sided cells 
in a typical environment of ordinary cells. 
Their embedding involves four distinct length scales, 
varying with four different powers of $n$. 
This has confirmed, among several other things, the very elongated shape 
of the first-neighbor cells. 

No particular effort was made to optimize the code.
Our total investment of computer time on a recent model PC 
was limited to a few hundred hours and
allowed us to obtain $p_n$ to at least
four or five significant decimals for the set of $n$ values
listed in the tables.
We have also not attempted to provide any `best fits' to the numerical curves,
as we have no reason to believe that there exist 
simple analytic expressions that fit the data 
within our error bars over their full range.

The Monte Carlo work of this paper became possible only after 
initial analytic progress \cite{HJHletter05,HJHart05}. 
We believe that it will in return spur 
further analytic investigation. One branch of study
may concern the nature of the
asymptotic expansions uncovered here.
Another one may deal with
correlations between a cell and its second, 
third, and higher topological neighbors, which are a recurrent theme in
the theory and applications of Voronoi cells. 
 
%%%%%%%%%%%%%%%%%%%%%%%%%%%%%%%%%%%%%%%%%%%%%%%%%%%%%%%%%%%%%%%%%%%%%%%%%%%%%

\section*{Acknowledgments}
The author thanks Daniela de Oliveira Maionchi for providing him with
the computer program that draws the Voronoi cells of a given set of particles.
He thanks Pierre Calka for discussions and for indicating to him
Ref.\,\cite{HayenQuine00b}.

%%%%%%%%%%%%%%%%%%%%%%%%%%%%%%%%%%%%%%%%%%%%%%%%%%%%%%%%%%%%%%%%%%%%%%%%%%%%%%
%%%%%%%%%%%%%%%%%%%%%%%%%%%%%%%%%%%%%%%%%%%%%%%%%%%%%%%%%%%%%%%%%%%%%%%%%%%%%%

\appendix

\section{Theory}
\label{apptheory}

We present here the extension of earlier work that opens the way
to the numerical simulations of this work.
We consider uniformly and independently distributed particles in the 
plane. Let a particle be placed in
the origin and let $n$ other particles
occupy the positions $2\bR_1, 2\bR_2,\ldots,2\bR_n$.
The sidedness probability $p_n$ of the cell containing the origin
may then be written 
as a $2n$-fold integral on the midpoint 
coordinates \cite{DI84,ID89,Calka03a,Calka03b},
\begin{equation}
{p}_n=\frac{1}{n!}
\int \dd\bR_1 \ldots \dd\bR_n\,\,
\chi(\bR_1,\ldots,\bR_n)\,\, 
\ee^{-{\cal A}(\bR_1,\ldots,\bR_n)}.
\label{exprR2pn}
\end{equation}
%%% [*2701 `history' of prefactors; 2951,2901] 
Here the indicator function $\chi$ 
is equal to unity (or to zero) on the domain of phase space where
the perpendicular bisectors of the vectors $2\bR_m$, for $m=1,2,\ldots,n$,
define an $n$-sided (or a fewer-sided) cell around the origin;
and ${\cal A}$ is the two-dimensional volume of the area
that should be void of particles if this cell is not to be intersected by any
of the bisectors of the position vectors of the remaining particles.
Explicit expressions for ${\cal A}$ and ${\chi}$ are given in
Refs.\,\cite{Calka03b,HJHart05}.

%%%%%%%%%%%%%%%%%%%%%%%%%%%%%%%%%%%%%%%%%%%%%%%%%%%%%%%%%%%%%%%%%%%%%%%%%%%%%

\subsection{Starting point}
\label{secnotation}

After one integrates over a common
radial scale, expression (\ref{exprR2pn})
takes the form \cite{HJHart05} of an integral on  the angle $\beta_1$
and on the sets of angles $\xi=\{\xi_m\}$
and $\eta=\{\eta_m\}$, 
\bea
{p}_n &=& 
\frac{(n-1)!}{2n}\int_{-\pi/2}^{\pi/2}\!\!\dd\beta_1
\int\!\!\dd\xi\dd\eta\,\,
\delta\big(\sum_{m=1}^n\xi_m-2\pi\big)\,
\delta\big(\sum_{m=1}^n\eta_m-2\pi\big) \nonumber\\[2mm]
&&\times\,
\frac{\delta ( \beta_1-\beta_* )}{G'(\xi,\eta;\beta_*)}\,
\Big[\prod_{m=1}^n \rho_m^2 T_m^{\phantom{2}} \Big]
[\pi(1+n^{-1}V)]^{-n},
\label{exprpn}
\eea
%%% [* I.2.48]
where $G$ and $\beta_*$ are as defined in the main text 
[Eqs.\,(\ref{defG}) and (\ref{nospiralcond})]
and the derivative $G'=\dd G/\dd\beta_1$ is given explicitly by
(\ref{Gprime}) and (\ref{exprbgtilde}) of Appendix \ref{appeqnG0}. 
The definitions of the new symbols occurring in (\ref{exprpn})
follow below.

First of all, the symbol $\int\!\dd\xi\dd\eta$ in (\ref{exprpn})
is shorthand for the nested integrations 
%%% [* I.2.49]
\bea
\int\dd\xi\dd\eta &=&
\int_0^{\pi/2+\beta_{_1}}\!\!\dd\xi_1 
\int_0^{\pi/2+\gamma_{_1}}\!\!\dd\eta_1
\int_0^{\pi/2+\beta_{_2}}\!\!\dd\xi_2 \ldots \nonumber\\[1mm]
&& \phantom{xxxxxxx}
\ldots\int_0^{\pi/2+\gamma_{_{n-1}}}\!\!\dd\eta_{n-1} 
\int_0^{\pi/2+\beta_{_n}}\!\!\dd\xi_n\,\,     
\int_0^{2\pi}\dd\eta_n\,.\phantom{xx}
\label{defintxieta}
\eea
The notation is hybrid; the variables
$\gamma_1,\beta_2,\gamma_3,\ldots,\beta_n$ 
occurring here
should be viewed as functions of $\xi$, $\eta$, and the `angle of rotation'
$\beta_1$. They are given by 
%%% [* I.2.16] 
\bea
\beta_m &=& \phantom{-}\beta_*(\xi,\eta) - 
\sum_{\ell=1}^{m-1}(\xi_{\ell}-\eta_\ell),
\nonumber\\
\gamma_m &=& -\beta_*(\xi,\eta) +
\sum_{\ell=1}^{m-1}(\xi_{\ell}-\eta_{\ell}) + \xi_m\,, 
\qquad m=1,\ldots,n,
\label{inversebgxy}
\eea
where $\sum_{\ell=1}^0$ denotes an empty sum.
Next, the $T_m$ and $\rho_m$ are functions of the $\gamma_m$ and 
$\beta_m$ given by 
%%% [* I.2.23; 2.33; 2.30]
%%% [* 3655]
\beq
T_m=\frac{\sin(\beta_m+\gamma_m)}{\cos^2\beta_m}, \qquad m=1,\ldots,n,
\label{defTm}
\eeq 
\beq
\rho_m=\frac{\cos\gamma_m\cos\gamma_{m-1}\ldots\cos\gamma_1}
{\cos\beta_m\cos\beta_{m-1}\ldots\cos\beta_1}\, \rho_n\,,
\qquad m=1,\ldots,n-1,
\label{condcosbcosg}
\eeq
and the condition
\beq
n^{-1}\summ\rho_m=1.
\label{sumrulerho}
\eeq
Finally $V$ is given by
%%% [* 1104,1204;2078 coeff in ${\cal A}$] 
%%% [* I.2.41]
\bea
V &=& 
\frac{n}{4\pi}\summ \rho_m^2 \big[ \tan\gamma_m-\gamma_m 
+\tan\beta_{m+1}-\beta_{m+1} \nonumber\\
&& \phantom{xxxxxxxxx} + \gamma_{m}\tan^2\gamma_{m}
   + \beta_{m+1}\tan^2\beta_{m+1} \big] \nonumber\\
&& +\, \frac{n}{2\pi}
   \summ (\rho_m^2 - 1)(\gamma_m + \beta_{m+1}). 
\label{exprVbetagamma}
\eea
The factor $n$ 
included in its definition makes that, typically, $V$ is of order 
$n^0$ as $n\to\infty$.
This completes the definition of the multiple integral
(\ref{exprpn}) for $p_n$.

%%%%%%%%%%%%%%%%%%%%%%%%%%%%%%%%%%%%%%%%%%%%%%%%%%%%%%%%%%%%%%%%%%%%%%%%%%%%%

\subsection{Transformations} 
\label{sectransformation}

We now depart from the development of Ref.\,\cite{HJHart05}
and transform expression (\ref{exprpn}) as follows.
We integrate over $\beta_1$ 
and henceforth when writing $\beta_1$ it will be understood that 
it takes the value $\beta_1=\beta_*(\xi,\eta)$.
The integration requires that Eq.\,(\ref{nospiralcond}) have a 
solution. In Ref.\,\cite{HJHart05}
a unique solution was shown to exist 
perturbatively for large $n$; in Appendix B
of the present work we provide the demonstration for general $n$.
We furthermore replace the upper integration limits of the integrals
over the $\xi_m$ and $\eta_m$ by $\infty$ at the expense
of introducing Heaviside theta functions. Using that
$\xi_m-\beta_m=\gamma_m$ and $\eta_m-\gamma_m=\beta_{m+1}$ and introducing a
factor $\theta(\frac{\pi}{2}-\beta_1)$, which may be done for free 
we find that Eq.\,(\ref{exprpn}) may be converted into 
\bea
p_n &=& \frac{(n-1)!}{2n\pi^n} \int_0^{\infty} 
\dd\xi_1\,\xi_1\ldots\dd\xi_n\,\xi_n
\int_0^{\infty} \dd\eta_1\ldots\dd\eta_n \nonumber\\[2mm]
&&\times\,
\delta(\summ\xi_m-2\pi)\,\delta(\summ\eta_m-2\pi)\,\Theta\ee^{-{\mathbb V}}, 
\label{pnsym}
\eea 
in which
%%% [* I.2.54]
\beq
\ee^{-{\mathbb V}} = G'(\xi,\eta;\beta_*)^{-1}
\Big[ \prod_{m=1}^n \rho_m^2 T_m^{\phantom{2}} \xi_m^{-1} \Big]
(1+n^{-1}V)^{-n}
\label{defmathbbV}
\eeq
and
\beq
\Theta = \prodm\theta(\tfrac{\pi}{2}-\beta_m)
\prodm\theta(\tfrac{\pi}{2}-\gamma_m).
\label{defTheta}
\eeq
Expression (\ref{pnsym}) is more symmetric than 
(\ref{exprpn})-(\ref{defintxieta}).
Its integrand is a function 
exclusively of the $\xi_m$ and $\eta_m$.
We have purposefully included extra weights $\xi_m$
in the integrations in (\ref{pnsym}) and compensated for these by 
factors $\xi_m^{-1}$ in the product on $m$ in (\ref{defmathbbV}).
In this way we obtain property that $T_m\xi_m$ 
remains finite when $\xi_m \to 0$, which was desirable analytically
\cite{HJHletter05,HJHart05} and is also necessary numerically.

The same quantity ${\mathbb V}$ as defined in (\ref{defmathbbV})
was studied analytically in
Refs.\,\cite{HJHletter05,HJHart05}, where it was shown that for $n\to\infty$
it remains, typically, of order $n^{0}$.

One further rewriting is useful. We set
\beq
\xi_m = \alpha_{2m-1}+\alpha_{2m}, \qquad m=1,2,\ldots,n,
\eeq
and use that 
\beq
\int_0^\infty\!\dd\alpha_{2m-1}\dd\alpha_{2m}\,f(\alpha_{2m-1}+\alpha_{2m})
=\int_0^\infty\!\dd\xi_m\,\,\xi_m f(\xi_m)
\eeq
for any function $f(\xi_m)$. This converts (\ref{pnsym}) into
the final result
%%% [* PVN 4,5]
\beq
p_n\,=\,p_n^{(0)}\la\Theta\ee^{-{\mathbb V}}\ra,
\label{defav}
\eeq
where for any function $X$ of the angular variables the average $\la X \ra$
is defined by
\bea
\la X \ra &=& \frac{1}{p_n^{(0)}}
\int_0^\infty \dd\alpha_1\ldots\dd\alpha_{2n}
\int_0^\infty \dd\eta_1\ldots\dd\eta_n \nonumber\\
&&\times\,
\delta\big(\sum_{m=1}^{2n}\alpha_m-2\pi\big)\,
\delta\big(\summ\eta_m-2\pi\big)\,X.
\label{pnfin}
\eea
The normalization factor $p_n^{(0)}$ that appears here
is easily calculated as
%%% [* compare I.4.1]
\bea
p_n^{(0)} &=& \frac{(n-1)!}{2n\pi^n} \Big[
  \int_0^\infty \dd\alpha_1\ldots\dd\alpha_{2n}\,
  \delta\big(\sum_{m=1}^{2n}\alpha_m-2\pi\big) \Big] \nonumber\\
    && \times \Big[ \int_0^\infty \dd\eta_1\ldots\dd\eta_n\,
       \delta\big(\summ\eta_m-2\pi\big) \Big] \nonumber\\
&=& \frac{(n-1)!}{2n\pi^n}\times\frac{(2\pi)^{2n-1}}{(2n-1)!}
              \times\frac{(2\pi)^{n-1}}{(n-1)!} \nonumber\\
&=& \frac{(8\pi^2)^n}{4\pi^2(2n)!}\,,
\label{exprcalN}
\eea
which is (\ref{resultpn0C}). This way of arriving at $p_n^{(0)}$ is 
slightly
simpler than the original calculation of Ref.\,\cite{HJHart05}.
Expressions (\ref{defav})-(\ref{pnfin}) are new and are at the basis of the 
Monte Carlo simulation of this work. The integrals in (\ref{pnfin})
directly suggest step (i) of the algorithm of subsection (\ref{secpn}).

%%%%%%%%%%%%%%%%%%%%%%%%%%%%%%%%%%%%%%%%%%%%%%%%%%%%%%%%%%%%%%%%%%%%%%%%%%%%%%

\section{The equation $G=0$}
\label{appeqnG0}

We discuss here the function $G$ defined by
\beq
\ee^{2\pi G} = \frac{\cos\gamma_1\cos\gamma_2\ldots\cos\gamma_n}
{\cos\beta_1\cos\beta_2\ldots\cos\beta_n}.
\label{defGbis}
\eeq
The transformation to angular variables in Appendix \ref{apptheory} 
led to Eq.\,(\ref{defTheta}), {\it i.e.} to the upper limits of integration
$\beta_m, \gamma_m < \frac{\pi}{2}$. Since $\beta_m+\gamma_m=\xi_m$ and since
$\xi_m>0$, we have in fact that in the integral for $p_n$ the angles $\beta_m$
and $\gamma_m$ are restricted by
\beq
-\tfrac{\pi}{2} < \beta_m, \gamma_m < \tfrac{\pi}{2}\,.
\label{dombetagamma}
\eeq
Hence $G\to -\infty$ whenever any of the $\gamma_m$ 
tends to $\pm\frac{\pi}{2}$, and $G\to\infty$
whenever any of the $\beta_m$ tends to $\pm\frac{\pi}{2}$.
We set now
\bea
\beta_m  &=& \tilde{\beta}_m  + \beta_1, \nonumber\\
\gamma_m &=& \tilde{\gamma}_m - \beta_1, \qquad m=1,\ldots,n,
\label{defbgtilde}
\eea
where the $\tilde{\beta}_m$ and $\tilde{\gamma}_m$ are functions of the
$\xi_m$ and $\eta_m$ that may be read off by a comparison of 
Eqs.\,(\ref{defbgtilde}) and (\ref{inversebgxy}),
\bea
\tilde{\beta}_m &=& - \sum_{\ell=1}^{m-1}(\xi_{\ell}-\eta_\ell),
%, \qquad m=2,\ldots,n, 
\nonumber\\
\tilde{\gamma}_m &=& \phantom{-}
\sum_{\ell=1}^{m-1}(\xi_{\ell}-\eta_{\ell}) + \xi_m\,, 
\qquad m=1,\ldots,n.
\label{exprbgtilde}
\eea
where again $\sum_{\ell=1}^0$ denotes the empty sum.
Making all $\beta_1$ dependence explicit we get
\beq
\ee^{2\pi G}\,=\, 
\frac{\cos(\tilde{\gamma}_1-\beta_1)\cos(\tilde{\gamma}_2-\beta_1)\ldots
\cos(\tilde{\gamma}_n-\beta_1)}
{\cos(\tilde{\beta}_1+\beta_1)\cos(\tilde{\beta}_2+\beta_1)\ldots
\cos(\tilde{\beta_n}+\beta_1)}
\label{Gbeta1}
\eeq
which we wish to study as a function of the single variable $\beta_1$,
at fixed $(\xi,\eta)$.
Expression (\ref{Gbeta1}) shows that \,$\exp(2\pi G)$\, is positive on 
the interval 
\beq
 -\tfrac{\pi}{2}+\max_{1\leq m\leq n}\tilde{\gamma}_m < \beta_1
< \tfrac{\pi}{2}-\max_{1\leq m\leq n}\tilde{\beta}_m,
\label{intervalbeta1}
\eeq
provided this interval is not empty, that is, provided 
\beq
\max_{1\leq m\leq n}\tilde{\gamma}_m +
\max_{1\leq m\leq n}\tilde{\beta}_m  < \pi.
\label{condinterval}
\eeq
Because of the preceding discussion, $G$ approaches $-\infty$ and $\infty$
as $\beta_1$ approaches the left and right hand
end points of this interval, respectively.
To show that $G$ is actually monotonous in $\beta_1$ on
the interval (\ref{intervalbeta1}), it suffices to analyze the derivative
%%% [* I.6.6]
\beq
\frac{\dd G}{\dd \beta_1} = \frac{1}{2\pi}
\summ\big[ \tan(\tilde{\gamma}_m-\beta_1) + 
           \tan(\tilde{\beta}_m+\beta_1) \big].
\label{Gprime}
\eeq
Since $\tilde{\beta}_m+\tilde{\gamma}_m=\xi_m>0$, 
it follows that there are three cases, namely
(i) $\tilde{\beta}_m,\tilde{\gamma}_m>0$,\,\, 
(ii) $\tilde{\beta}_m>0,\, \tilde{\gamma}_m<0$, 
and (iii) $\tilde{\beta}_m<0,\, \tilde{\gamma}_m>0$.
By considering each of them separately one deduces that the summand in
Eq.\,(\ref{Gprime}) is always positive. It follows that $\dd G/\dd\beta_1>0$
and hence that $G=0$ has a unique solution $\beta_1=\beta_*(\xi,\eta)$
in the interval (\ref{intervalbeta1}).

Hence we have shown that the conditions $\xi_m,\eta_m>0$ and
$\beta_m,\gamma_m<\frac{\pi}{2}$ suffice for the equation $G=0$ to have a
unique solution $\beta_*$ in the physical interval (\ref{intervalbeta1}).
This condition involves, however, the angles $\beta_m$ and $\gamma_m$ which
are determined by the solution $\beta_*$. We would like to have a criterion
for the existence of a solution in terms of the sole sets $(\xi,\eta)$
given at the outset.
By retracking the solution method, we see that it is valid for all
$(\xi,\eta)$ as long as the `physical' interval (\ref{intervalbeta1}) is not
empty, that is, as long as Eq.\,(\ref{condinterval}) is satisfied.
When made explicit with the aid of (\ref{exprbgtilde}), 
Eq.\,(\ref{condinterval}) becomes condition 
(\ref{nospiraldomain}) of the main text.
\vspace{10mm}

%%%%%%%%%%%%%%%%%%%%%%%%%%%%%%%%%%%%%%%%%%%%%%%%%%%%%%%%%%%%%%%%%%%%%%%%%%%%%%


\begin{thebibliography}{10}
\bibitem{footnotec}
The ${\tPsi}_m$
are equal to the angles $\Psi_m$ of Fig.\,\ref{figdefangles} 
up to a common additive constant 
which drops out of Eq.\,(\ref{defetam}).
An analogous remark applies to the $\tPhi_m$.
\bibitem{footnotef}
A factor $4\lambda$ instead of $\lambda$ appears because
the particle positions were defined as $2\bR_m$ instead of $\bR_m$.
\bibitem{footnoteg}
For a figure comparing their data to our asymptotic result
we refer to Ref.\,\cite{HJHletter05}.
\end{thebibliography}
\end{document}